\documentclass[a4paper,fleqn]{cas-dc}

\usepackage[numbers]{natbib}

\usepackage{graphicx}
\usepackage{multirow}
\usepackage{amsmath,amssymb,amsfonts}
\usepackage{amsthm}
\usepackage{mathrsfs}
\usepackage{xcolor}
\usepackage{textcomp}
\usepackage{manyfoot}
\usepackage{booktabs}
\usepackage{algpseudocode}
\usepackage{listings}
\usepackage{balance}
\usepackage[ruled,vlined]{algorithm2e}

\begin{document}
	
	\let\WriteBookmarks\relax
	\def\floatpagepagefraction{1}
	\def\textpagefraction{.001}
	
	\shorttitle{Rethinking Few-Shot Image Fusion: Granular Ball Priors Enable General-Purpose Deep Fusion}
	
	\shortauthors{Deng et al.}
	
	\title [mode=title]{Rethinking Few-Shot Image Fusion: Granular Ball Priors Enable General-Purpose Deep Fusion}

\author[1]{Minjie Deng}[orcid=0009-0002-8132-7023]
\ead{minjiedeng@qq.com}

\author[1]{Yan Wei}
\cormark[1]
\ead{weiyancq@163.com}

\author[1]{An Wu}
\ead{2024110516047@stu.cqnu.edu.cn}

\author[1]{Yuncan Ouyang}
\ead{2023210516060@stu.cqnu.edu.cn}

\author[1]{Hao Zhai}
\ead{zhaihao@cqnu.edu.cn}

\author[1]{Qianyao Peng}
\ead{2024110516042@stu.cqnu.edu.cn}

\cortext[1]{Corresponding author}

	\begin{abstract}		
In image fusion tasks, the absence of real fused images as supervision signals poses significant challenges for supervised learning. Existing deep learning methods typically address this issue either by designing handcrafted priors or by relying on large-scale datasets to learn model parameters. Different from previous approaches, this paper introduces the concept of incomplete priors, which formally describe handcrafted priors at the algorithmic level and estimate their confidence. Based on this idea, we couple incomplete priors with the neural network through a sample-level adaptive loss function, enabling the network to learn and re-infer fusion rules under conditions that approximate the real fusion process.
To generate incomplete priors, we propose a Granular Ball Pixel Computation (GBPC) algorithm based on the principles of granular computing. The algorithm models fused-image pixels as information units, estimating pixel weights at a fine-grained level while statistically evaluating prior reliability at a coarse-grained level. This design enables the algorithm to perceive cross-modal discrepancies and perform adaptive inference.
Experimental results demonstrate that even under few-shot conditions, a lightweight neural network can still learn effective fusion rules by training only on image patches extracted from ten image pairs. Extensive experiments across multiple fusion tasks and datasets further show that the proposed method achieves superior performance in both visual quality and model compactness. The code is available at: https://github.com/DMinjie/GBFF
.
		
	\end{abstract}
	

	
	
	\maketitle
\section{Introduction}

Due to the inherent limitations of sensors, images captured by a single sensor often have restricted representational capacity and cannot effectively capture all the information within the scene. The goal of image fusion is to combine the information from images acquired by different sensors, integrating multiple modalities into a single image that contains richer and more comprehensive information. For example, visible-light images captured at night cannot record thermal sources in dark regions, whereas their fusion with infrared images can effectively enhance the representation of thermal information. Because the fused image integrates the advantages of multiple sensing devices, it has found broad applications in fields such as medical imaging, security surveillance, target recognition, and image segmentation.

In previous research, significant progress has been made in areas such as multi-focus image fusion, multi-exposure image fusion, infrared and visible image fusion, and medical image fusion. These methods can generally be divided into traditional approaches and deep learning-based approaches. Traditional methods typically decompose images using algorithmic operations at different scales or spatial domains to extract features for fusion\cite{vif_survey}. However, as practical requirements in real-world scenarios have grown, the design of these algorithms has become increasingly complex.
Consequently, deep learning methods have gradually become dominant. Among them, deep learning-based approaches can be categorized into convolutional neural network (CNN)-based methods, generative adversarial network (GAN)-based methods, Transformer-based methods, diffusion model-based methods, and hybrid approaches that integrate traditional algorithms with deep learning. These approaches have achieved remarkable results in the field\cite{mff_survey}.
Traditional methods provide priors that guide neural networks to form semi-manual fusion rules, which has been proven effective in enhancing the expressive power of fused images-for example, through wavelet transforms, multi-scale decompositions, and saliency-based techniques. However, previous hybrid methods that combine traditional algorithms with deep learning often relied on fixed loss functions to learn supervised images derived from traditional methods as complete priors, lacking the coupling and adaptability between algorithms and networks. As a result, a large number of training samples are still required for convergence, making few-shot training difficult to achieve.

To address the aforementioned challenges, we introduce concepts from granular computing theory and model the neural network learning process as a form of re-reasoning over uncertain information based on known information. Through this formulation, the neural network no longer needs to repeatedly model already-determined information; instead, it concentrates its learning capacity on uncertain regions, thereby reducing the dependence on large-scale training samples.
Specifically, this paper proposes the Granular Ball Pixel Computation (GBPC) algorithm. The proposed algorithm generates fusion priors through multi-granularity analysis. At the fine-grained level, adaptive granular balls are used as computational scales to calculate pixel-level weights, enabling initial pixel-level fusion. At the coarse-grained level, we adopt a fuzzy rough class formulation. By analyzing two similarity classes composed of meta-granular balls, the method distinguishes reliable regions from regions that require further reasoning. Among them, regions with high confidence are defined as the positive domain (POS), while regions requiring further inference are defined as the boundary domain (BND).
Since this prior is incomplete, the statistical information of POS and BND is utilized to dynamically adjust the loss function, enabling the neural network to adaptively regulate the learning process according to sample characteristics and perform re-reasoning based on the prior. Within this framework, the incomplete prior can be regarded as a degraded image annotated with regional confidence labels, while the neural network is responsible for completing the mapping from the original input images to the real fused image. Because the prior information is not complete, the network can effectively avoid the overfitting problem caused by fully determined priors.
During training, the input images are randomly cropped. Owing to the adaptive nature of the proposed algorithm, distinctive priors can still be formed even from image fragments, thereby simulating complex real-world environments. This property enables the model to maintain strong generalization ability under few-shot training conditions.The main contributions of this paper are summarized as follows:
\begin{itemize}
	\item{To the best of our knowledge, this work represents the first attempt to introduce granular computing into general-purpose multimodal image fusion, covering infrared–visible image fusion, multi-exposure image fusion, multi-focus image fusion, and medical image fusion, and establishes a unified fusion framework that provides a new theoretical perspective for fusion algorithm design.}
	\item{We propose the concept of incomplete priors and design the Granular Ball Pixel Computation (GBPC) algorithm. The method directly represents co-located pixel features through meta-granular balls and performs granular computation based on feature similarity without relying on explicit spatial partitioning. Furthermore, meta-granular balls are jointly analyzed at both coarse-grained and fine-grained levels.}
	\item{A deep coupling mechanism between incomplete priors and neural networks is established to achieve sample-wise adaptive learning. By simplifying the information that the neural network needs to learn through incomplete priors, the learning objective is transformed from modeling the source data distribution to performing re-reasoning based on the prior of the current sample, thereby enabling few-shot training.}
	\item{Extensive qualitative and quantitative comparisons with state-of-the-art (SOTA) methods are conducted across multiple fusion tasks. Despite being trained on only ten images, the proposed method still demonstrates clear advantages in both fusion quality and deployment cost, verifying its strong generalization capability and practical effectiveness.}
\end{itemize}
The structure of the remaining sections is organized as follows. In Section 2, we introduce existing image fusion methods and the fundamentals of granular computing. Section 3 provides a detailed description of the proposed fusion framework and algorithmic implementation. In Section 4, we present comparative analyses between our method and existing approaches. Finally, Section 5 offers further theoretical discussions and extensive experimental validations.
\section{Related Work}
In this section, we will introduce the methods and developments of image fusion, as well as the theory and applications of granular computing.
\subsection{Image Fusion}

Image fusion requires methods capable of extracting features from different modalities and highlighting the effective information contributed by each modality after fusion, thereby fully representing the real scene \cite{james2014medical11}. In previous studies, researchers mainly adopted approaches such as multi-scale transformations, subspace analysis, and sparse representation. As application scenarios became increasingly complex, some researchers started combining multiple techniques to further enhance information extraction capability \cite{zhang2021image34}. Early studies utilized the Laplacian Pyramid (LP) to decompose images into multiple frequency bands, followed by the introduction of the Dual-Tree Complex Wavelet Transform (DWCWT), which extracts effective information via different filtering operations. However, wavelet-based methods exhibit poor edge extraction performance, motivating the development of curvelet transform (Curvelet), which captures edge structures more effectively. To enhance the representation of structural information in fusion, Guided Filtering (GFF) was also incorporated into image fusion \cite{li2013image12}. In addition, Liu et al. combined sparse matrix representation with traditional transform methods and achieved promising results in extracting effective fusion information \cite{liu2015general35}.

However, in real-world applications, as the usage demands increase, image processing algorithms often suffer from limited generalization capability. This leads to increasingly complex algorithmic designs, posing significant challenges for practical deployment \cite{zhang2021image34}. With the rise of deep learning, Li et al. introduced it into the field of image fusion for the first time \cite{li2018densefuse14}, making it gradually become a mainstream research direction.
Based on model architectures, the earliest encoder-decoder network, DeepFuse, demonstrated strong detail extraction capability for infrared and visible images \cite{ram2017deepfuse13}. Subsequently, IFCNN leveraged pre-trained networks to extract features and achieved general-purpose fusion within a CNN framework, while U2Fusion proposed a saliency-based loss function to implicitly guide cross-modality fusion. Unlike CNN-based methods, GAN-based architectures use a generator-discriminator mechanism to evaluate the fused outputs. FusionGAN \cite{ma2019fusiongan15} was the first to apply GANs to image fusion, and Ma et al. further proposed DDcGAN \cite{ma2020ddcgan16}, where multiple discriminators were utilized to improve fusion accuracy. For Transformer-based architectures, long-range dependency modeling is exploited by decomposing images into tokens for processing \cite{han2022survey36}. On this basis, Ma et al. designed cross-domain attention using Swin Transformer to enhance fusion quality \cite{ma2022swinfusion1}. Furthermore, diffusion models have also been introduced into fusion tasks. Chen et al. developed a multi-exposure diffusion framework by designing flexible fusion strategies \cite{UTFusion}.

In addition to model-driven methods, approaches free from network constraints have also been proposed. By leveraging deep priors to perform sample-free fusion directly on images, ZMFF treats multi-focus masks as denoising priors to accomplish multi-focus fusion \cite{ZMFF}, while ZVIR directly utilizes the network structure to extract infrared and visible features, thus achieving image fusion without data dependence \cite{zvir}. Bai et al. further proposed Refusion, which performs unsupervised fusion by meta-learning the loss function \cite{refusion}. In addition, Wang et al. introduced a fusion method that combines vision-language models with image data \cite{wang2025multi}.
With the development of deep learning, hybrid approaches that integrate traditional algorithms with neural networks have also emerged. For example, MMIF-INET integrates wavelet decomposition with invertible networks for medical image fusion \cite{MMIF}, while Wang et al. later proposed a semi-supervised fusion framework that combines Laplacian pyramid transformers with neural networks \cite{wang2024general}. Moreover, Kang et al. extracted fusion features within Euclidean space and developed SMLNet based on this strategy \cite{kang2025smlnet}.

Although these methods have achieved remarkable success, existing trainable fusion approaches inevitably rely on large-scale datasets to progressively update model parameters, while sample-free methods still suffer from high computational cost, making practical deployment challenging. Moreover, deep learning methods that depend on traditional priors often struggle to adaptively determine the fusion scale; even when prior information is incorporated, a substantial amount of external data is still required to constrain the learning of network parameters. These limitations collectively hinder the rapid deployment of image fusion techniques, especially in few-shot and real-world application scenarios.
\subsection{Granular Ball Computing	}

Granular computing organizes fine-grained raw data into higher-level information granules and performs analysis, reasoning, and computation across multiple granularity levels.
Since real-world information often exhibits hierarchical structures as well as uncertainty and ambiguity, analyzing data at different levels of granularity helps reveal relationships among information granules.
In the spatial Granular Ball model, a granular ball is typically represented by a centroid and a radius~\cite{cheng2025gbmod32,xia2023gbrs31}.
Building upon these foundations, researchers proposed the GBNRS method, which integrates the granulation mechanism with approximation theory to handle continuous data more effectively~\cite{xia2020gbnrs30}.
Subsequently, granular ball clusters were introduced into clustering tasks to model relationships among instances, leading to the Ball k-means algorithm that significantly reduces spatial complexity during clustering~\cite{xia2020ball33}.
Granular Ball computing has demonstrated strong potential in data analysis and uncertainty modeling~\cite{xie2023efficient}.
Later, Xia {et al.} employed granular balls to construct graph representations by partitioning spatial regions for graph neural network learning\cite{xia2025adaptive}.
Building upon this idea, Chen {et al.} further introduced granular-ball-based block representations and applied graph neural networks to multi-focus image fusion\cite{chen2025gbfusion}.

However, these approaches typically partition images into spatial blocks using granular balls, effectively forming superpixels.
Due to the lack of a unified granulation criterion, such spatial partition strategies are difficult to generalize to multi-modal image fusion tasks.
To establish a unified information granulation standard, we propose to represent the features carried by pixels in image fusion using meta-granular balls, which form the universe of discourse.
Granular balls evolve along the corresponding feature axis, enabling the model to capture local equivalence at a fine granularity and similarity at a coarse granularity.
This design results in a general framework that does not rely on explicit spatial partitioning.
Furthermore, considering the inherent limitations of algorithmic priors, we estimate the reliability of the generated priors and introduce the concept of an incomplete prior.

\section{Proposed Methods}

\subsection{Framework}
\begin{figure*}[htbp]
	\centering
	\includegraphics[width=0.9\textwidth]{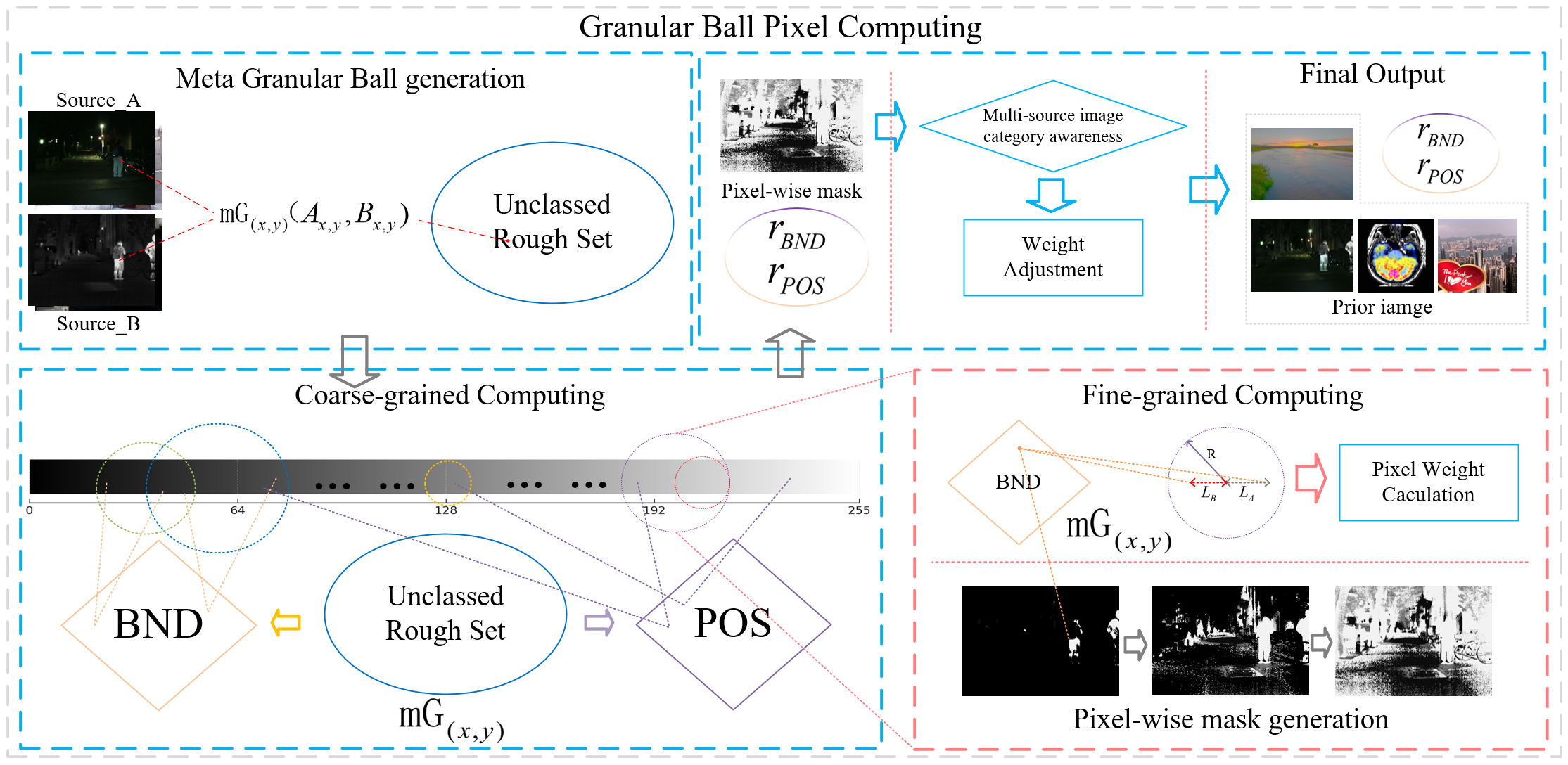} 
	\caption{Illustration of the Granular Ball Pixel Computing (GBPC) framework and visualization of the prior formation process.}
	\label{struct_1}
\end{figure*}
In our proposed framework, the training phase can be divided into two main parts: one involves generating incomplete priors using the Granular Ball Pixel Computation algorithm, and the other involves the neural network performing re-inference on the basis of these incomplete priors to obtain the final fused image. To verify the effectiveness of the proposed priors, we deliberately employ a simple CNN architecture without introducing any additional regularization modules such as attention mechanisms. This ensures that all information within the network is transmitted and regulated solely through the loss function.
For each pair of training input images, the pixel pairs are first granulated to form meta-granular balls. The Granular Ball Pixel Computation algorithm then mines the internal information of each meta-granular ball and performs fusion to generate a pseudo-supervised image. Due to the algorithm-s limited inference capacity, the resulting prior image contains incomplete information compared with the real fused image. Therefore, we use rough set theory to describe the confidence of this prior, which allows us to regulate the loss function dynamically-achieving adaptive coupling between the algorithm and the network.
In summary, each pair of input images corresponds to an independent prior and a corresponding loss function. Under a few-shot setting, by cropping the images into multiple patches, a large number of independent and diverse samples can be generated, enabling the neural network to perform inference based on incomplete priors and thus achieve generalization to real-world scenarios.
\subsection{Definitions and Concepts}
\begin{figure*}[htbp]
	\centering
	\includegraphics[width=0.9\textwidth]{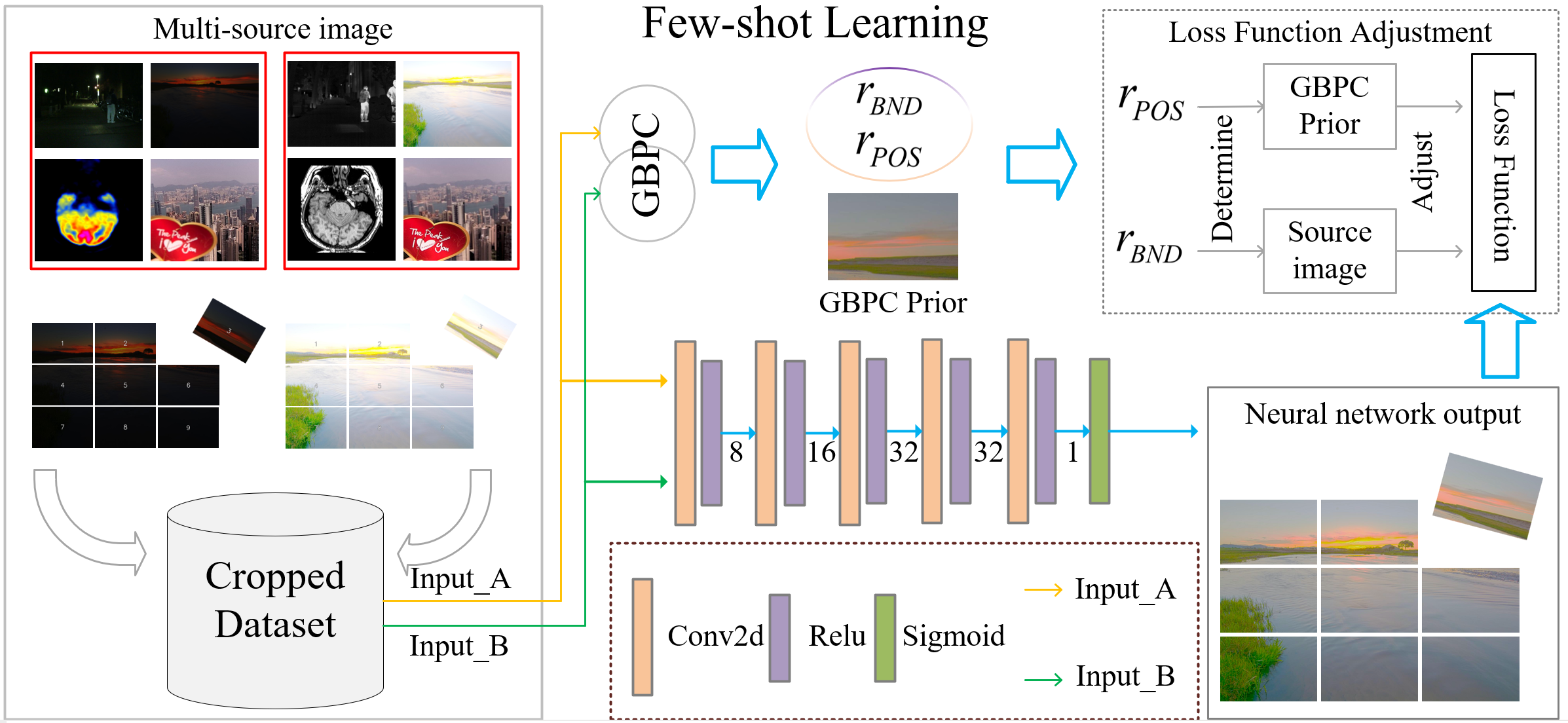} 
	\caption{Illustration of the network architecture and the processing steps involved during the learning procedure.}
	\label{struct_2}
\end{figure*}
In a pair of images to be fused, images from different modalities carry different types of information. The differences between pixels at the same spatial location can therefore serve as key cues for evaluating information saliency. Accordingly, we perform modeling in the YCbCr color space and project the two-dimensional luminance of different modalities into a one-dimensional space, using granular balls as the analysis scale to enable the extraction of pixel pairs with equivalent contributions by exploiting the self-recursive distribution of pixels and their differences within the image pair.
The proposed framework employs three complementary forms of granularity: meta-granular balls for information representation, granular balls for adaptive scale discovery, and coarse-grained semantic regions for modeling cross-modal discrepancy.

First, for a pair of input images, granular-ball modeling is performed. The definition of the meta-granular ball is given in Definition 1.

\noindent\textbf{Definition 1. Meta-Granular Ball.} 
A Meta-Granular Ball, denoted as \(mG(x, y)\), represents the paired attributes of two corresponding pixels at coordinate \((x, y)\) from image \(A\) and image \(B\). In general, these attributes can be multi-dimensional feature representations extracted from the two modalities. Formally, it can be expressed as
\begin{equation}
	mG(x,y) = (\mathbf{F}_A(x,y),\, \mathbf{F}_B(x,y)),
\end{equation}
where \(\mathbf{F}_A(x,y)\) and \(\mathbf{F}_B(x,y)\) denote the feature vectors of the pixels at coordinate \((x,y)\) in images \(A\) and \(B\), respectively. Since the luminance component in the Y channel has been shown to contain sufficient information for image fusion\cite{U2,DDB}, the feature vectors are instantiated as pixel brightness values for computational simplicity,
\(
\mathbf{F}_A(x, y)=L_A(x, y), \quad
\mathbf{F}_B(x, y)=L_B(x, y).
\)

This meta-granular ball serves as the {elementary unit} for subsequent granular-ball construction, 
providing the basis for local information interaction between the two modalities.
For a pair of input images, an independent universe of discourse $U$ is defined as
\begin{equation}
	U = \{ mG(x,y) \mid (x,y) \in \Omega \}
\end{equation}
where $\Omega $ denotes the set of image coordinates.

Since different image pairs exist in distinct information environments, granular balls are introduced as the analysis scale. By using granular balls, the internal information differences within meta-granular balls can be adaptively characterized and processed according to the corresponding information environment. The spatial definition of a granular ball is given in Definition 2.

\noindent \textbf{Definition 2. Granular Ball} 
We define a granular ball $\mathcal{G}(\mu, r)$ in the brightness space as:
\begin{align}
	\mathcal{G}(\mu, r) &= \left\{ x \in \mathbb{R} \,\middle|\, |x - \mu| \leq r \right\}, & \mathcal{G}(\mu, r) \subseteq [0, 255]
\end{align}
where $r$ denotes the radius of the granular ball, $\mu$ represents the center, 
$\mathbb{R}$ is the real number domain, and $x$ corresponds to any element within the meta-granular ball.
The motion of granular balls in space, such as sliding, expansion, and splitting, gradually forms the corresponding decision regions.

\subsection{Granular Ball Pixel Computing}
\subsubsection{Construction of the Granular-Ball Scale}
\begin{figure}[t]  
	\centering
	\includegraphics[width=0.85\linewidth]{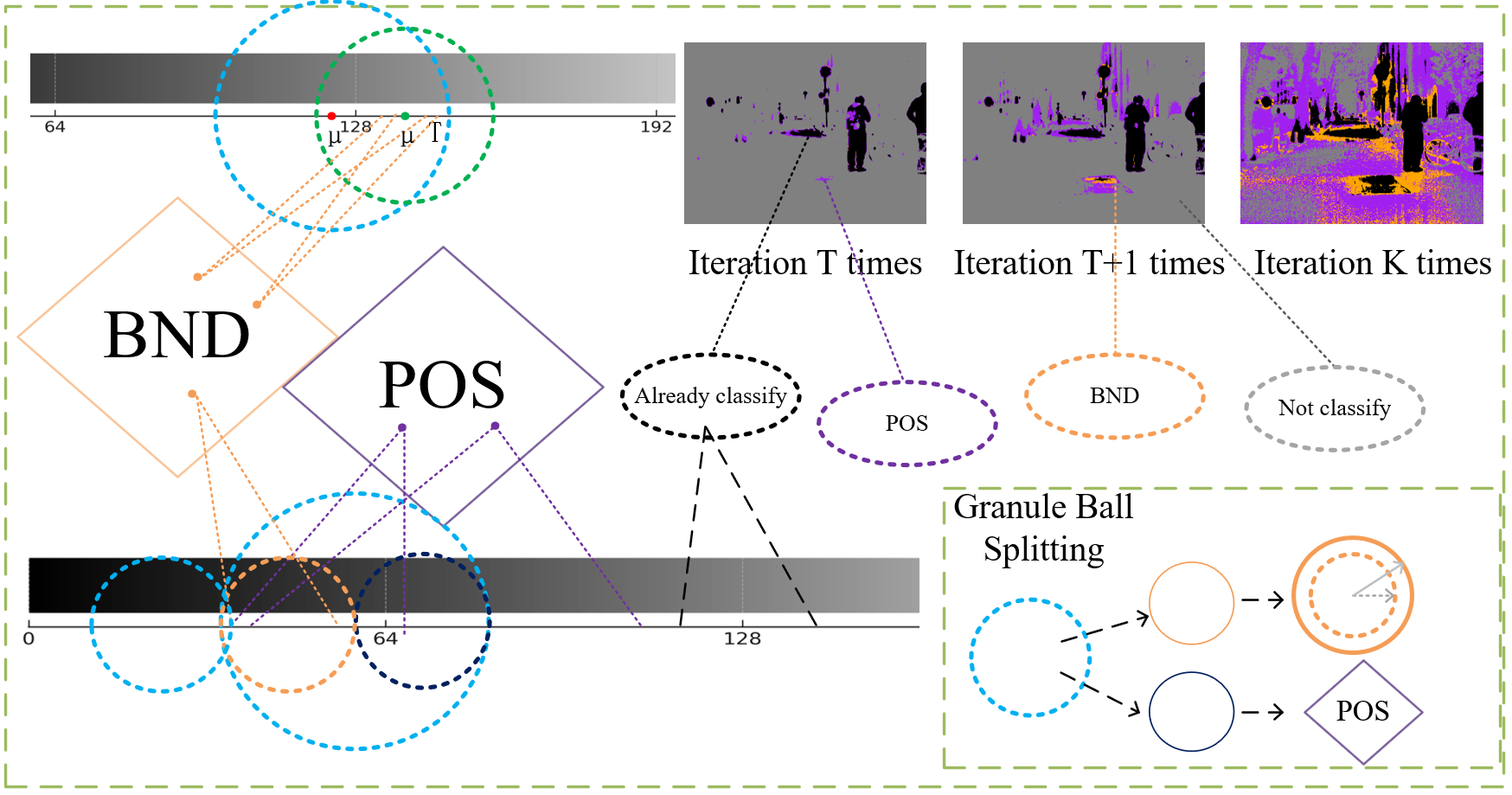}
	\caption{Visualization of the meta-granular ball capturing process in Granular Ball Pixel Computing for clearer understanding.}
	\label{kd}
\end{figure}
Granular balls are initialized at the boundaries with an initial radius of zero, ensuring that every meta-granular ball can be captured during the traversal process. The initial radius and the sliding step size of the granular balls are predefined, while the splitting limit of the granular balls is also determined by hyperparameters. The iterative process of granular balls consists of three operations: sliding, expansion, and splitting. Among them, the expansion and splitting behaviors are controlled by the predefined hyperparameters, whereas the sliding is governed by the distribution of meta-granular balls, enabling the granular balls to exhibit adaptive behavior. The sliding operation is triggered when no meta-granular ball exists within the interval of length $\Delta m$ to the left of the current granular-ball boundary, while at least one meta-granular ball exists within the interval of length $\Delta m$ to the right boundary. In this case, the granular ball slides rightward by $\Delta m$, which is formulated as
\begin{align}
	\mathcal{G}_{t+1}(\mu, r) = \mathcal{G}_{t}(\mu+\Delta m, r).
\end{align}
where $\Delta m$ is determined by the meta-granular ball closest to the current boundary, and $t$ denotes the state of the granular ball at the $t$-th iteration.

Since granular balls ignore the spatial distribution information of pixels, while the luminance distribution within a single modality exhibits spatial continuity, this continuity is exploited to extract granular balls with similar attributes, and the differences between modalities are then used for decision-making. Accordingly, the expansion of granular balls is designed based on the continuity of the spatial distribution, aiming to capture meta-granular balls with similar attributes within a single modality as much as possible. The expansion of granular balls can be described as:
\begin{align}
	\mathcal{G}_{t+1}(\mu, r) = \mathcal{G}_{t}(\mu, r+\Delta d)
\end{align}
where $\Delta d$ denotes the incremental step length of radius expansion in each iteration.

When the granular ball expands to its maximum limit, a splitting process is triggered. 
Therefore, when \( r \geq k\Delta d \), the granular ball undergoes division as follows:

\begin{align}
	\mathcal{G}_{t+1(L)}(\mu, r) &= \mathcal{G}_{t}(\mu - r/2, r/2) , \\
	\mathcal{G}_{t+1(R)}(\mu, r) &= \mathcal{G}_{t}(\mu + r/2, r/2)
\end{align}
where the auxiliary granular ball $\mathcal{G}_{t+1(L)}(\mu,r)$ is introduced for determining the membership of elements in a meta-granular ball, while $\mathcal{G}_{t+1(R)}(\mu,r)$ serves as the granular ball for the next iteration.
Based on the above representations, the expansion and splitting processes of granular balls will organize the meta-granular balls into decision domains, which are used to determine the positive and boundary regions in the following section.

\subsubsection{Decision Domains Induced by Granular Balls}
The granular-ball expansion and splitting processes induce decision domains over the meta-granular ball space.The decision region is denoted as $R_i$, which is defined as:
Let $U^{(1)} = U$ denote the initial universe of discourse, and let $U^{(i)}$ denote the residual universe before the $i$-th iteration.
At iteration $i$, the granular ball generated by the current evolution state induces a decision domain $R_i$ over the residual universe, defined as
\begin{equation}
R_i = 
\{ mG \in U^{(i)} \mid 
mG \subseteq G_t(\mu,r)
\}
\end{equation}
After the assignment, all meta-granular balls in $R_i$ are removed from the residual universe:
\begin{equation}
	U^{(i+1)} = U^{(i)} \setminus R_i.
\end{equation}
Thus, each meta-granular ball is assigned exactly once, and the set of decision domains $\{R_i\}_{i=1}^{T}$ forms a partition of the original universe $U$.
The decision domains induced by granular-ball evolution satisfy the following properties:
\begin{equation}
	\bigcup_{i=1}^{T} R_i = U,
\end{equation}
\begin{equation}
	R_i \cap R_j = \emptyset, \quad i \neq j,
\end{equation}
which indicates that each meta-granular ball belongs to only one decision domain.These domains naturally induce a partition over the meta-granular ball space, which provides the basis for defining granular decision regions.

Depending on whether the meta-granular balls remain indistinguishable or become separable under the current granular-ball scale, the induced decision domains can be categorized into boundary domains (BND) and positive domains (POS).
When the elements within a meta-granular ball remain indistinguishable at the current granular-ball scale, the captured meta-granular balls do not exhibit significant discriminative characteristics for fusion. In this case, the corresponding decision domain is regarded as a boundary domain (BND), since the similar attributes of the elements make it difficult to directly determine reliable fusion decisions.
In contrast, when the granular ball reaches the splitting condition, the elements within certain meta-granular balls become separable under the child granular-ball scales. This separability reflects significant cross-modal discrepancies, indicating that these meta-granular balls contain informative structures that can serve as reliable prior knowledge for the fusion task. Consequently, the corresponding decision domains are categorized into the positive domain (POS).
Therefore, the POS and BND regions characterize the discriminative capability of meta-granular balls under the current granular-ball scale and provide the basis for the subsequent prior reliability estimation.

For any meta-granular ball $mG(x,y)$ associated with the granular ball $G_t(\mu,r)$ corresponding to the decision domain $R_t$, the membership of its elements is determined by the type of the decision domain. Formally,

\noindent\textbf{Definition 3. Boundary Domain.}
Let $mG(x,y)$ be a meta-granular ball associated with the current granular ball $G_t(\mu,r)$.
If both elements of $mG(x,y)$ are covered by the same granular ball at the current scale and
satisfy the fuzzy similarity relation, then $mG(x,y)$ is regarded as indistinguishable at this scale.
In this case, the corresponding decision domain is classified as a boundary domain.
Formally, the boundary domain is defined as
\begin{equation}
	\mathcal{R}_{BND} =
	\left\{
	R_t \;\middle|\;
	\{L_A(x,y),\,L_B(x,y)\}\in G_t(\mu,r)\;
	\right\}.
\end{equation}
Then the boundary region is given by
\begin{equation}
	BND = \bigcup_{R_t\in\mathcal{R}_{BND}} R_t .
\end{equation}

\noindent\textbf{Definition 4. Positive Domain.}
Let $mG(x,y)$ be a meta-granular ball associated with the current granular ball $G_t(\mu,r)$.
If the current granular ball satisfies the splitting criterion and the two elements of $mG(x,y)$
become separable under the child granular-ball scales, then $mG(x,y)$ is considered to exhibit
significant cross-modal discrepancy. In this case, the corresponding decision domain is classified
as a positive domain.
Formally, let $\mathcal{G}_{t+1(L)}(\mu_L,r_L)$ and $\mathcal{G}_{t+1(R)}(\mu_R,r_R)$ denote
the two child granular balls generated by splitting $G_t(\mu,r)$. The positive domain is defined as
\begin{equation}
	\mathcal{R}_{POS} =
	\left\{
	R_t \;\middle|\;
	\begin{aligned}
		&r \ge k\Delta d,\\
		&\exists\, G' \in \{\mathcal{G}_{t+1(L)},\mathcal{G}_{t+1(R)}\} \\
		&\text{s.t. } |\{L_A(x,y),L_B(x,y)\}\cap G'| = 1,\\
	\end{aligned}
	\right\}.
\end{equation}
Then the positive region is given by
\begin{equation}
	POS = \bigcup_{R_t\in\mathcal{R}_{POS}} R_t .
\end{equation}

The proposed granular-ball framework naturally induces two levels of
equivalence relations ,which correspond to coarse-grained semantic consistency
and fine-grained computational consistency.

\noindent\textbf{Definition 5. Fine-Grained Equivalence.}
Let \(R_t\) denote a decision domain induced by the
granular ball \(G_t(\mu,r)\).
A fine-grained equivalence relation \(\sim_f\) is defined as
\begin{equation}
	mG_p \sim_f mG_q
	\iff
	\exists i \quad
	mG_p \in R_i \ \land\ mG_q \in R_i .
\end{equation}

In this case, all meta-granular balls captured by the same
granular ball \(G_t(\mu,r)\) form an equivalence class,
which shares the same analysis scale and statistical context.

\noindent\textbf{Definition 6. Coarse-Grained Semantic Consistency.}
Let $POS$ and $BND$ denote the positive and boundary regions induced by
granular-ball evolution over the meta-granular ball universe $U$.
Two meta-granular balls $mG_p$ and $mG_q$ are said to satisfy
\emph{coarse-grained semantic consistency}, denoted by
$mG_p \sim_c mG_q$,
if and only if they belong to the same semantic region, i.e.,
\begin{equation}
	\begin{aligned}
		mG_p \sim_c mG_q
		\iff {} & (mG_p,mG_q \in POS) \\
		& \text{or} \\
		& (mG_p,mG_q \in BND).
	\end{aligned}
\end{equation}
Since POS and BND form a partition of the universe U,
the relation $\sim_c$ satisfies reflexivity,
symmetry, and transitivity, thus constituting
a coarse-grained equivalence relation.This relation indicates that the corresponding meta-granular balls share
consistent semantic attributes at the coarse granularity level.

\subsubsection{Pixel Weight Computing}
For any meta-granular ball $mG(x,y)$ that belongs to a decision domain,
there exists a unique granular-ball scale $G(\mu,r)$ associated with it.
Within the coverage range of the corresponding granular ball,
the computation scheme for the meta-granular balls in the boundary domain
is defined as follows:

\begin{align}
	D = \max\{L_A(x,y),\, L_B(x,y)\}, 
\end{align}

\begin{align}
	L = \max\left( \left|D - (\mu - r)\right|,\, \left|D - (\mu + r)\right| \right),
\end{align}

\begin{align}
	w(x,y) = \frac{L}{2r},
\end{align}

\begin{align}
	w_A(x,y) &=
	\begin{cases}
		w(x,y), & L_A(x,y)> L_B(x,y),\\[4pt]
		1 - w(x,y), & L_A(x,y) \le L_B(x,y),
	\end{cases}
\end{align}

\begin{align}
	w_B(x,y) &= 1 - w_A(x,y)
\end{align}
where $w_A(x,y)$ and $w_B(x,y)$ denote the weights of pixels from images $A$ and $B$ within the current meta-granular ball, respectively.

For the meta-granular balls in the positive domain, their discrepancy
exceeds the scale defined by the granular ball. Therefore, the splitting
condition of the granular ball is adopted as the scale criterion.
Assume that in the next iteration step the sliding process will capture
the other element of the meta-granular ball and it will lie on the boundary.
Under this assumption, the corresponding weight must satisfy
\begin{equation}
	w(x,y) \geq 1 - \frac{\Delta d}{2k\Delta d + \Delta d}.
\end{equation}
To reduce computational complexity, the weight $w(x,y)$ is therefore set as $w(x,y) = 1 - \frac{\Delta d}{2k\Delta d + \Delta d}$.
Subsequently, $w_A(x,y)$ and $w_B(x,y)$ are computed according to equations (21) and (22).

\subsubsection{Modality Perception}
The statistical distribution of POS domains provides a coarse indicator of cross-modal discrepancy within the current patch. 
During the iterative evolution of granular balls, each decision attribute domain is associated with a corresponding decision domain. This association enables a coarse-grained estimation of the attributes of meta-granular balls within the current universe. 
For images from different modalities, particularly in multi-exposure image fusion, certain fusion regions exhibit significant intensity discrepancies. In such regions, the proportion of meta-granular balls belonging to the positive domain (POS) tends to increase significantly. This statistical property provides a useful indicator for identifying regions with strong cross-modal discrepancies.
To quantify this phenomenon, we define the observed ratios of meta-granular balls belonging to the positive and boundary domains as
\begin{align}
	\triangle r_{POS} &= \frac{\mathrm{count}(POS)}{N},\\
	\triangle r_{BND} &= \frac{\mathrm{count}(BND)}{N},
\end{align}
where $\mathrm{count}(POS)$ and $\mathrm{count}(BND)$ denote the numbers of meta-granular balls assigned to the POS and BND domains, respectively, and $N$ represents the total number of meta-granular balls.
To regulate the influence of highly discrepant exposure regions, the observed POS ratio $\triangle r_{POS}$ is compared with a predefined threshold $M$. The threshold $M$ is determined according to the statistical distribution of POS ratios observed across different fusion tasks. Empirical analysis shows that multi-exposure fusion often produces significantly higher POS ratios due to strong intensity discrepancies between input images. Therefore, the threshold is set as $M=0.95$ in this work. The influence of $M$ will be further analyzed in the ablation study section.
The optimization coefficients are determined as
\begin{equation}
	r_{POS}, r_{BND} =
	\begin{cases}
		(\triangle r_{POS}, \triangle r_{BND}), & \triangle r_{POS} < M, \\[6pt]
		(0.5,\,0.5), & \triangle r_{POS} \ge M .
	\end{cases}
\end{equation}

When $\triangle r_{POS}$ exceeds the threshold $M$, the algorithm identifies fusion regions with significant cross-modal discrepancies. In this case, the pixel weights $w(x,y)$ corresponding to meta-granular balls belonging to the POS domain are reset to $0.5$, enforcing equal contributions from both modalities, while meta-granular balls in the boundary domain remain unchanged. This adjustment suppresses the dominance of highly discrepant exposure regions during optimization.
It is worth noting that, since image patches are used as the feature library, the behavior observed at the local scale manifests differently at the global image level. Specifically, the generated prior image tends to suppress over-exposed regions while preserving relatively well-exposed areas. Therefore, the distribution of POS domains provides a natural statistical indicator for detecting exposure anomalies in multi-exposure image fusion.
Finally, the prior image can be computed as
\begin{equation}
	\begin{aligned}
		I_{\text{prior}}(x,y) =&\; w_A(x,y)\cdot L_A(x,y) \\
		&+ w_B(x,y)\cdot L_B(x,y).
	\end{aligned}
\end{equation}

Based on the above steps, the overall process of granular ball pixel computing is 
summarized in Algorithm~1.
\begin{algorithm}[htbp]
	\caption{Granular Ball Pixel Computing}
	\label{alg:gbpc}
	\KwIn{Two input images $A$ and $B$}
	\KwOut{Prior image $I_{\mathrm{prior}}$, domain confidences $r_{\mathrm{POS}}$ and $r_{\mathrm{BND}}$}
	
	Construct the meta-granular ball set $U$ and initialize $POS,BND \leftarrow \emptyset$\;
	\While{$U \neq \emptyset$}{
		Initialize a granular ball at the boundary with $r\leftarrow 0$\;
		Expand the granular ball by $\Delta d$ and slide it by $\Delta m$ if needed\;
		Determine the induced decision domain $R_t$\;
		\eIf{$r < k\Delta d$}{
			Assign $R_t$ to $BND$ and compute weights for $mG(x,y)\in R_t$\;
		}{
			Split the granular ball, assign $R_t$ to $POS$, and compute weights for $mG(x,y)\in R_t$\;
		}
		$U \leftarrow U \setminus R_t$\;
	}
	Compute $\triangle r_{\mathrm{POS}}$ and $\triangle r_{\mathrm{BND}}$\;
	\If{$\triangle r_{\mathrm{POS}} \ge M$}{
		Reset the $w_A(x,y)=w_B(x,y)$ in $POS$ to $0.5$\;
	}
	Perform modality perception to obtain $r_{\mathrm{POS}}$ and $r_{\mathrm{BND}}$\;
	Compute $I_{\mathrm{prior}}(x,y)=w_A(x,y)L_A(x,y)+w_B(x,y)L_B(x,y)$\;
	\Return $I_{\mathrm{prior}}, r_{\mathrm{POS}}, r_{\mathrm{BND}}$\;
\end{algorithm}
\subsection{Hyperparameter Sensitivity Analysis}
\begin{figure}[t]  
	\centering
	\includegraphics[width=0.85\linewidth]{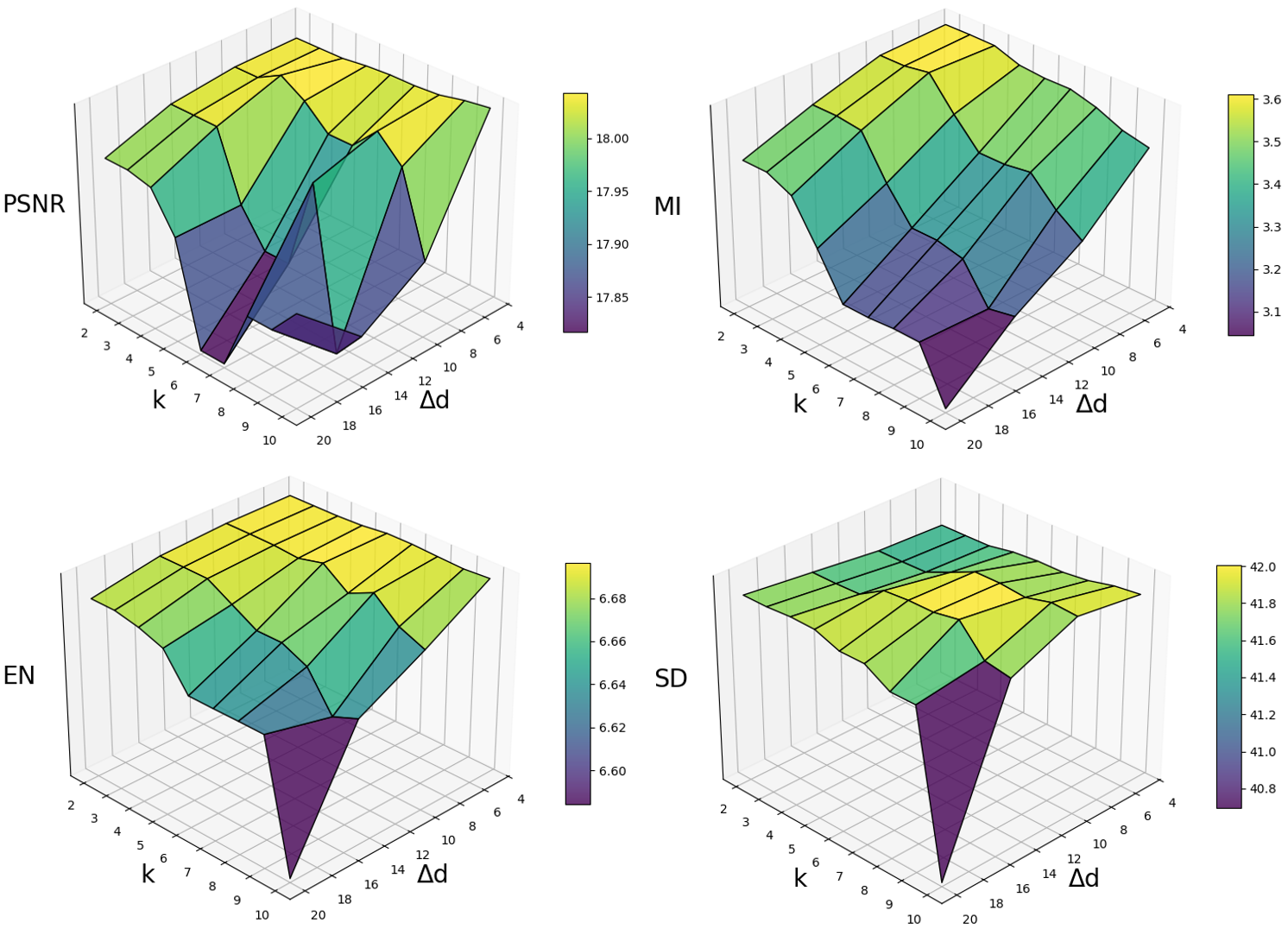}
	\caption{Objective evaluation variations of the prior image under different values of $\displaystyle k$ and $\displaystyle \Delta d$.}
	\label{kd}
\end{figure}
In Granular Ball Pixel Computing (GBPC), the parameters $k$ and $\Delta d$ jointly determine the perceptual domain and the splitting limit of each granular ball, thereby defining its overall scale. 
To ensure the robustness of the parameter configuration, we evaluated the quality of the prior images generated by the algorithm. 
At the algorithmic level, higher-quality prior images serve as the benchmark for determining the optimal values of $k$ and $\Delta d$, and several evaluation metrics were measured, as shown in Fig.\ref{kd}. 
Comprehensive analysis of all evaluation metrics indicates that when $k = 6$ and $\Delta d = 10$, 
the prior images achieve the best overall performance across multiple indicators, 
demonstrating superior representational quality and algorithmic stability.
\subsection{Few-shot learning}
In our framework, unlike previous approaches, we integrate algorithmic inference and neural network inference to enable the network to learn a complete set of rules. The neural network is responsible for re-inferring the uncertain information in GBPC based on the existing prior. The entire process is carried out through a loss function, whose design dynamically adapts according to the prior image. In GBPC, the prior image represents an approximate inference. When calculating the elements of the meta-granular balls located in the BND, the edge information tends to become blurred. Therefore, the neural network re-infers the missing edge information from the source images based on the structural brightness of the prior image. The loss function is composed of three parts, and its overall formulation can be described as follows:
\begin{figure}[t]  
	\centering
	\includegraphics[width=0.85\linewidth]{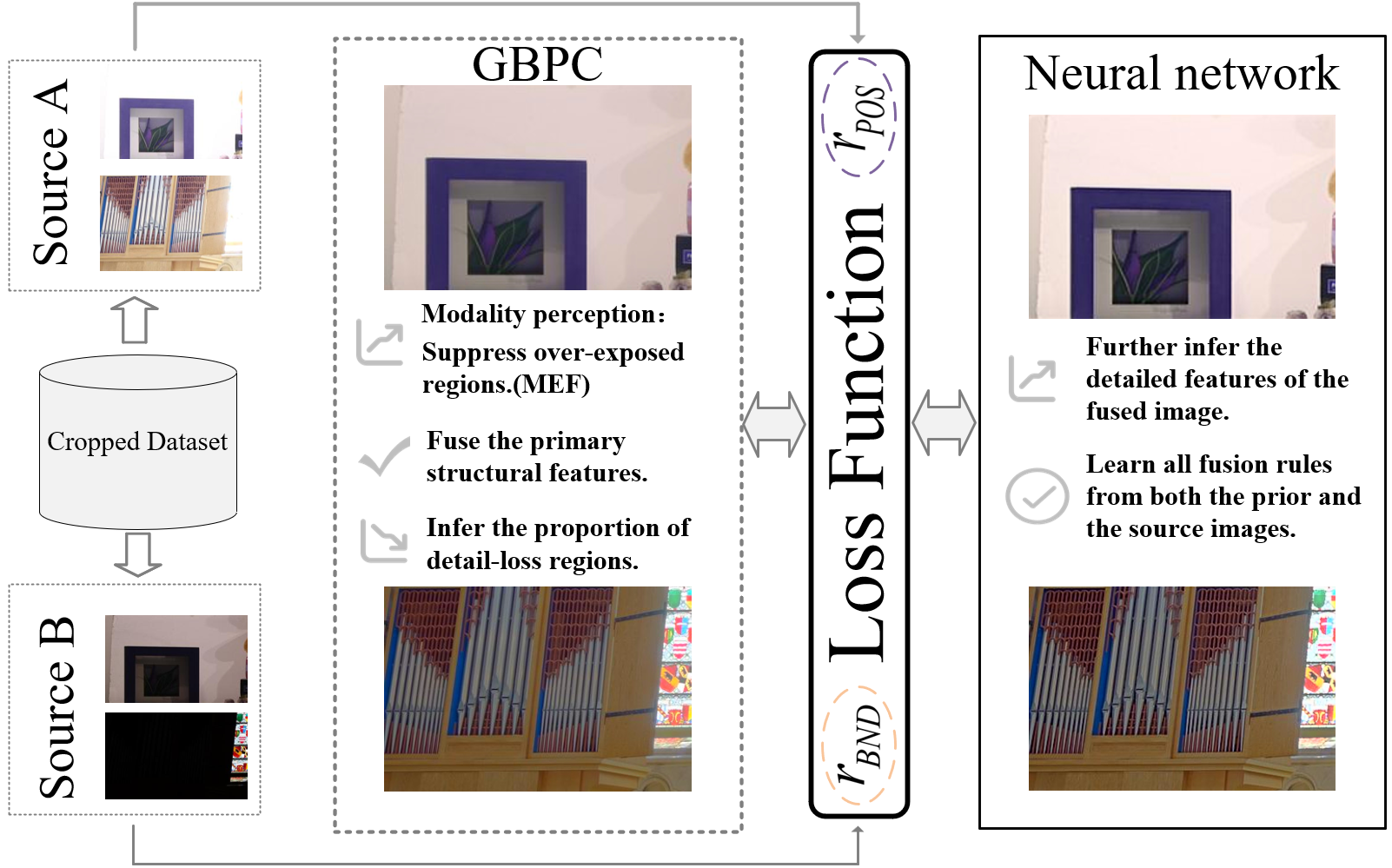}
	\caption{The prior is obtained from the fragment library using Granular Ball Pixel Computing, and the loss function is dynamically adjusted according to $r_{\mathrm{POS}}$ and $r_{\mathrm{BND}}$.}
	\label{crop}
\end{figure}
\begin{equation}
	L_{\text{total}} = L_{\text{SSIM}} + L_{\text{POS}} + L_{\text{BND}},
\end{equation}
where $L_{\text{SSIM}}$ transfers structural features from the prior image, 
$L_{\text{POS}}$ conveys reliable edge information within the prior, 
and $L_{\text{BND}}$ guides the network to infer weak regions of the prior 
based on feature extraction from the source images.The specific formulation of the loss function is defined as follows:
\begin{equation}
	L_{\text{SSIM}} = 1 - \text{SSIM}(Out, Prior),
\end{equation}
where $Out$ denotes the fused output image generated by the network, and $Prior$ represents the prior image obtained through the Granular Ball Pixel Computing (GBPC).
\begin{equation}
	L_{\text{POS}} = r_\text{POS} \times \mathcal{L}_1(\text{Sobel}(Out), \text{Sobel}(Prior)),
\end{equation}
where $r_{\text{POS}}$ denotes the proportion of the credible region in the prior image, which is also derived from GBPC. 
$\mathcal{L}_1(\cdot)$ represents the L1 loss function used to measure pixel-wise differences, 
and $\text{Sobel}(\cdot)$ denotes the Sobel operator used for gradient extraction.
\begin{equation}
	\begin{aligned}
		L_{\text{BND}} = &~r_\text{BND} \times \mathcal{L}_1(\text{Sobel}(Out), (\text{Sobel}(A) + \text{Sobel}(B))) \\
		&+ \mathcal{L}_1(\text{Lap}(Out), (\text{Lap}(A) + \text{Lap}(B))).
	\end{aligned}	
\end{equation}
where $r_{\text{BND}}$ represents the region with weaker edge responses in the prior image. 
This loss function assists the neural network in inferring correct regions from the source images $A$ and $B$, 
where $\text{Lap}$ denotes the Laplacian operator (second-order derivative) that helps to guide the reasoning of the first-order gradient extracted by the Sobel operator.
Specifically, the Sobel and Laplacian operators mentioned above are as follows:
\begin{equation}
	G_x =
	\begin{bmatrix}
		-1 & 0 & 1 \\
		-2 & 0 & 2 \\
		-1 & 0 & 1
	\end{bmatrix},
	\quad
	G_y =
	\begin{bmatrix}
		-1 & -2 & -1 \\
		0 & 0 & 0 \\
		1 & 2 & 1
	\end{bmatrix},
\end{equation}
\begin{equation}
	\text{Sobel}(I) = \sqrt{G_x^2*I + G_y^2*I}, \quad 
\end{equation}
where $G_x$ and $G_y$ extract horizontal and vertical gradient features, respectively.

\begin{equation}
	\text{Lap}(I) =
	\begin{bmatrix}
		0 & 1 & 0 \\
		1 & -4 & 1 \\
		0 & 1 & 0
	\end{bmatrix}*I,
\end{equation}

\section{Experimental results and analysis.}
\subsection{Experimental settings}
To perform thorough validation, publicly available datasets were selected according to the requirements of each task.
For the MEF (multi-exposure image fusion) task, we used the MEFB dataset\cite{zhang2021benchmarking}.
For the MFF (multi-focus image fusion) task, we used the Lytro dataset and the MFI-WHU dataset\cite{zhang2021mff}.
For the VIF (infrared and visible image fusion) task, we used the M3FD, MSRS, RoadScene, and TNO datasets\cite{U2}.
For the MIF (medical image fusion) task, we used the Harvard (PET-MRI) dataset.

Regarding the composition of the training set, for the MEF task, we selected 10 images from the MEFB dataset as training samples.
For the MFF task, we selected 10 images from the Lytro dataset.
For the VIF task,we selected 5 image pairs from TNO dataset and 5 from MSRS dataset.
For the MIF task, we selected 10 images from the Harvard (PET-MRI) dataset.
These training images are not included in the test set.

All selected images with resolutions higher than $640 \times 480$ were first resized to $640 \times 480$. 
The resized images were then cropped into $128 \times 128$ patches with a stride of 64 to construct the training dataset, 
and random flipping was applied during preprocessing. 
During training, the number of epochs was set to 100.
To ensure fairness of comparison, all experiments were conducted on a computing platform equipped with an RTX 3090 GPU and an Intel i5-13600KF CPU.

\subsection{Comparative Analysis}
\begin{table*}[t]
	\centering
	\caption{Comparison of different fusion methods on MEF, MFF, MIF, and VIF tasks.}
	\label{method}
	\resizebox{\textwidth}{!}{
		\begin{tabular}{lcccccccccccccccc}
			\toprule
			\textbf{Method} & EAT\cite{EAT} & SMAT\cite{SMAT} & UTFusion\cite{UTFusion} & VDMU\cite{VDMUF} & CCSR\cite{CCSR} & DMAnet\cite{DMA} & GRFusion\cite{GRFusion} & ZMFF\cite{ZMFF} & TextFusion\cite{textfusion} & Coconet\cite{coconet} & EMMA\cite{EMMA} & MMIF\cite{MMIF} & DDBFusion\cite{DDB} & GifFusion\cite{GIF} & U2Fusion\cite{U2} & Ours \\
			\midrule
			\textbf{Years} 
			& 2025 & 2024 & 2025 & 2024 & 2025 & 2025 & 2023 & 2024 & 2025 & 2024 & 2024 & 2025 & 2025 & 2025 & 2020 & -- \\
			\midrule
			\textbf{MEF} 
			& \checkmark & \checkmark & \checkmark & \checkmark & $\times$ & $\times$ & $\times$ & $\times$ & $\times$ & $\times$ & $\times$ & $\times$ & \checkmark & \checkmark & \checkmark & \checkmark \\
			\textbf{MFF} 
			& $\times$ & $\times$ & $\times$ & $\times$ & \checkmark & \checkmark & \checkmark & \checkmark & \checkmark & \checkmark & \checkmark & \checkmark & \checkmark & \checkmark & \checkmark & \checkmark \\
			\textbf{MIF} 
			& $\times$ & $\times$ & $\times$ & \checkmark & $\times$ & $\times$ & $\times$ & $\times$ & $\times$ & \checkmark & \checkmark & \checkmark & \checkmark & \checkmark & \checkmark & \checkmark \\
			\textbf{VIF} 
			& $\times$ & $\times$ & $\times$ & $\times$ & $\times$ & $\times$ & $\times$ & $\times$ & $\times$ & $\times$ & $\times$ & $\times$ & \checkmark & \checkmark & \checkmark & \checkmark \\
			\midrule
			\textbf{Params (M)}
			& \underline{0.016} & 1.233 & 556.805 & 31.773 & 0.039 & 4.076 & 7.71 & 4.66 & 0.074 & 6.845 & 1.52 & 0.749 & 10.921 & 7.71 & 0.659 & \textbf{0.015} \\
			\textbf{FLOPs (G)}
			& \underline{1.522} & 94.909 & 60659 & 188149 & 3.935 & 87.001 & 25.375 & 42.73 & 68.1 & 63.61 & 13.56 & 10.86 & 369.842 & 25.375 & 66.1 & \textbf{1.502} \\
			\textbf{Times (ms)}
			& 5.843 & 5.386 & 88060 & 20304 & 52.481 & 15.152 & 12.75 & 9100 & 3.116 & 6.512 & 17.1 & 6.17 & 80.49 & 12.75 & \underline{3.016} & \textbf{0.333} \\
			\bottomrule
		\end{tabular}
	}
\end{table*}
To verify the effectiveness of the proposed method, we selected three general-purpose fusion approaches and, for each specific task, additionally included four task-oriented fusion methods for comparison.
For clarity and ease of observation, the tasks associated with each comparative method are summarized in Table \ref{method}.
To ensure a fair comparison, all experiments were conducted on the same computing 
platform, and the implementations were reproduced using the model weights recommended 
by the original authors.

To conduct quantitative evaluation, we adopted several widely used objective metrics, including MI, EN, PSNR, VIF, SCD, $Q_g$, Qab, CC, and AG.\cite{mff_survey}
Mutual Information (MI) measures how much information the fused image shares with the source images.
Entropy (EN) evaluates the amount of information or the richness of details contained in the fused result.
Peak Signal-to-Noise Ratio (PSNR) reflects the fidelity of the fused image with respect to a reference image.
Visual Information Fidelity (VIF) assesses the visual quality of the fused image based on human visual perception.
The Sum of Correlation Difference (SCD) quantifies how well the fused image preserves complementary information from the source images.
The gradient-based quality metric ($Q_g$) evaluates the preservation of edges and structural details.
The Qab metric (Qab) measures the amount of edge information transferred from the source images to the fused output.
The Correlation Coefficient (CC) indicates the linear similarity between the fused image and each input image.
Finally, the Average Gradient (AG) reflects the sharpness and texture richness of the fused result.
For clearer observation, we annotate the objective evaluation metrics by marking 
the best scores in {bold} and the second-best scores with {underlines}.

\subsubsection{Multi-exposure image fusion}
\begin{table*}[t]
	\centering
	\caption{Quantitative comparison on MEFB and Harvard (PET--MRI) datasets.}
	\label{MEF_lab}
	\resizebox{\textwidth}{!}{
		\begin{tabular}{lccccc lccccc}
			\toprule
			\multirow{2}{*}{Method} &
			\multicolumn{5}{c}{MEFB} &
			\multirow{2}{*}{Method} &
			\multicolumn{5}{c}{Harvard (PET--MRI)} \\
			\cmidrule(lr){2-6} \cmidrule(lr){8-12}
			& EN & MI & PSNR & CC & Qab
			&  & MI & SD & PSNR & AG & Qab \\
			\midrule
			EAT
			& 6.819 & \textbf{6.207} & 10.378 & 0.880 & 0.598
			& VDMU & \textbf{3.467} & 60.339 & 15.468 & 10.614 & 0.413 \\
			
			SAMT-MEF
			& 6.949 & 4.792 & \underline{10.635} & \underline{0.881} & \underline{0.698}
			& COCONet & 2.570 & 62.769 & 11.852 & 15.437 & 0.479 \\
			
			UltraFusion
			& \textbf{7.308} & 2.652 & 7.738 & 0.596 & 0.226
			& EMMA & 2.991 & 85.119 & \underline{16.002} & 17.512 & 0.696 \\
			
			VDMU
			& 6.752 & 4.982 & 10.594 & 0.880 & 0.480
			& MMIF & 3.009 & \textbf{87.985} & 15.573 & \underline{19.504} & \underline{0.715} \\
			
			BBDFusion
			& \underline{7.131} & 4.604 & 10.412 & 0.875 & 0.458
			& BBDFusion & 2.927 & 69.864 & 15.121 & 13.051 & 0.374 \\
			
			GIFNet
			& 6.796 & 4.288 & 10.618 & 0.858 & 0.291
			& GIFNet & 2.629 & 79.149 & 14.448 & 15.930 & 0.457 \\
			
			U2Fusion
			& 6.739 & 2.744 & 8.326 & 0.616 & 0.222
			& U2Fusion & 2.726 & 62.874 & 15.400 & 8.397 & 0.287 \\
			
			\textbf{Ours}
			& 6.938 & \underline{5.116} & \textbf{10.638} & \textbf{0.883} & \textbf{0.708}
			& \textbf{Ours} & \underline{3.022} & \underline{85.355} & \textbf{16.010} & \textbf{20.282} & \textbf{0.717} \\
			\bottomrule
		\end{tabular}
	}
\end{table*}

\begin{figure*}[htbp]  
	\centering
	\includegraphics[width=0.9\linewidth]{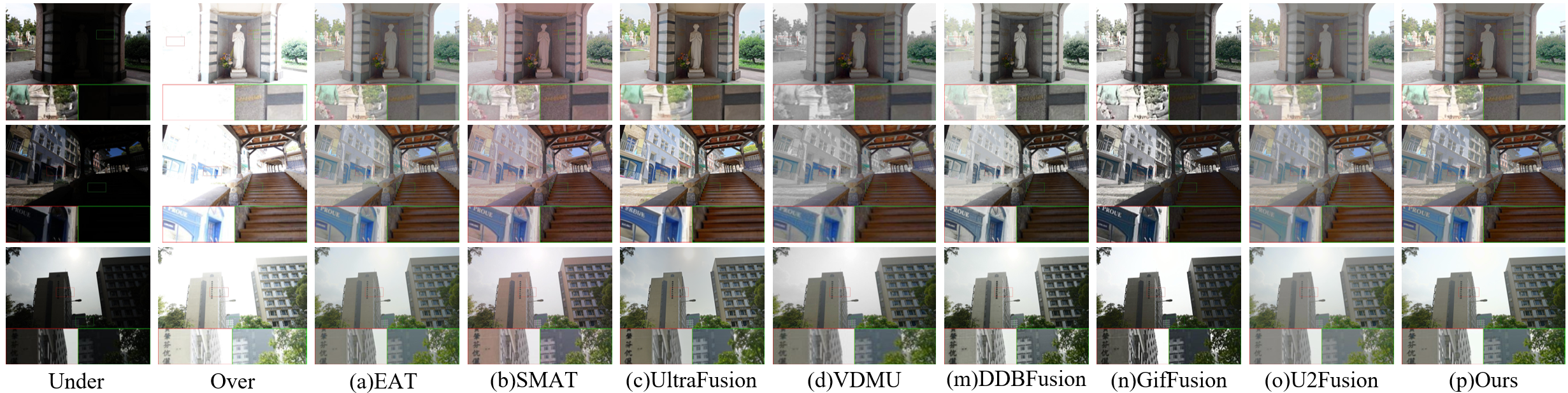}
	\caption{Qualitative comparison of fused results obtained by different methods on the MEF task.}
	\label{MEF}
\end{figure*}

For convenient comparison, we present the fusion results of different methods in Fig.\ref{MEF}, where the detailed regions are further enlarged for visual examination.
Compared with other approaches, our method effectively suppresses detail loss caused by overexposure while maintaining stable brightness without noticeable degradation.
Among the displayed methods, our approach also achieves clearer text rendering, while the diffusion-based method (c) UltraFusion, although better preserving color vividness, shows partial disappearance of characters.
As shown in Table \ref{MEF_lab}, our method achieves the best or second-best performance in MI, PSNR, CC, and Qab, indicating that it can preserve the original image information while maintaining superior overall image quality.
\subsubsection{Medical image fusion}
\begin{figure*}[htbp]  
	\centering
	\includegraphics[width=0.9\linewidth]{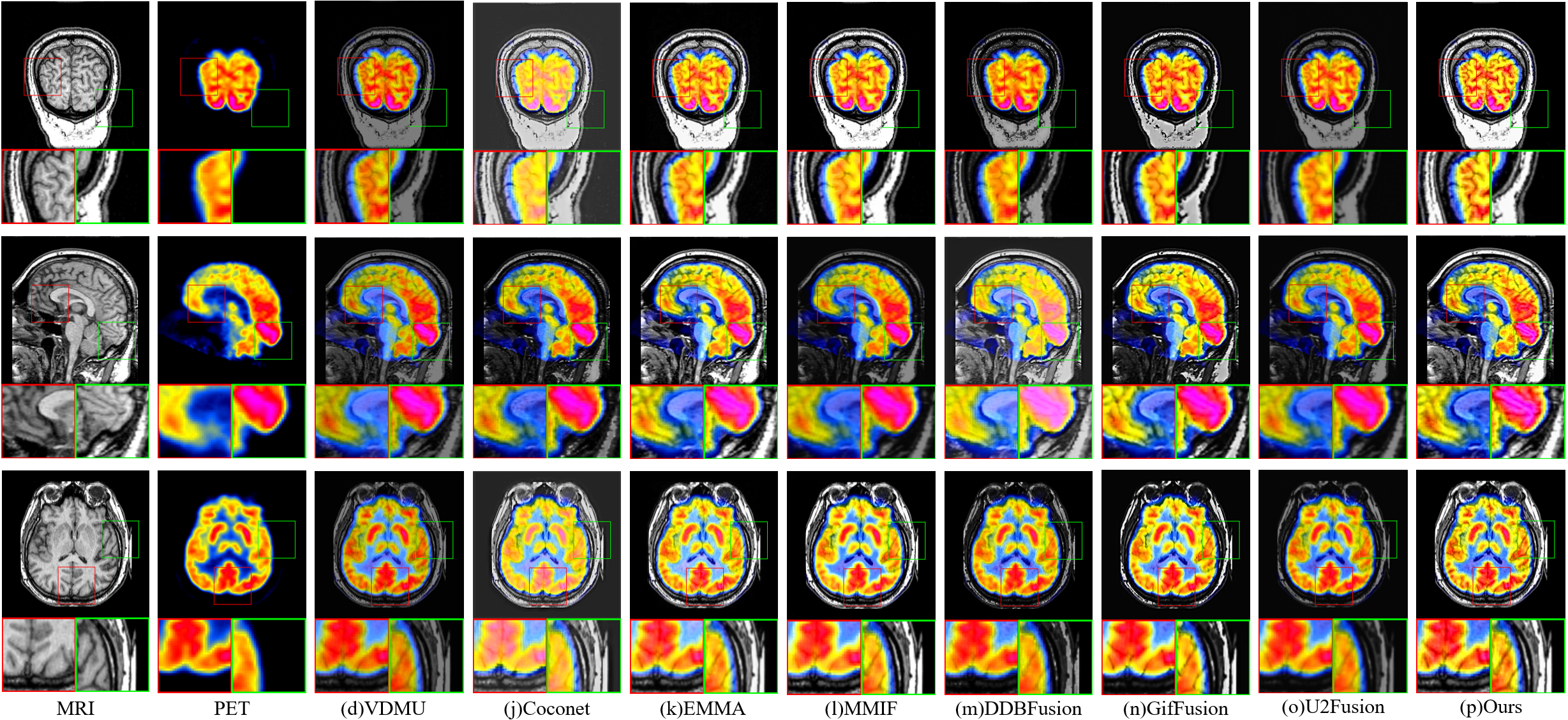}
	\caption{Qualitative comparison of fused results obtained by different methods on the MIF task.}
	\label{MIF}
\end{figure*}

Figure~\ref{MIF} shows the fusion results produced by different methods, 
and for clearer observation, the regions rich in details are further enlarged.
From the magnified regions of the fused images, we can observe that our method 
effectively preserves the color characteristics of the PET images while maintaining 
the complex structural details.
The edges in the MRI images are also well preserved during the fusion process, 
which verifies the effectiveness of our method.
As shown in Table~\ref{MEF_lab}, our method achieves the best or second-best 
performance across all objective metrics, demonstrating its overall effectiveness.
\subsubsection{Multi-focus image fusion}
\begin{figure*}[htbp]  
	\centering
	\includegraphics[width=0.9\linewidth]{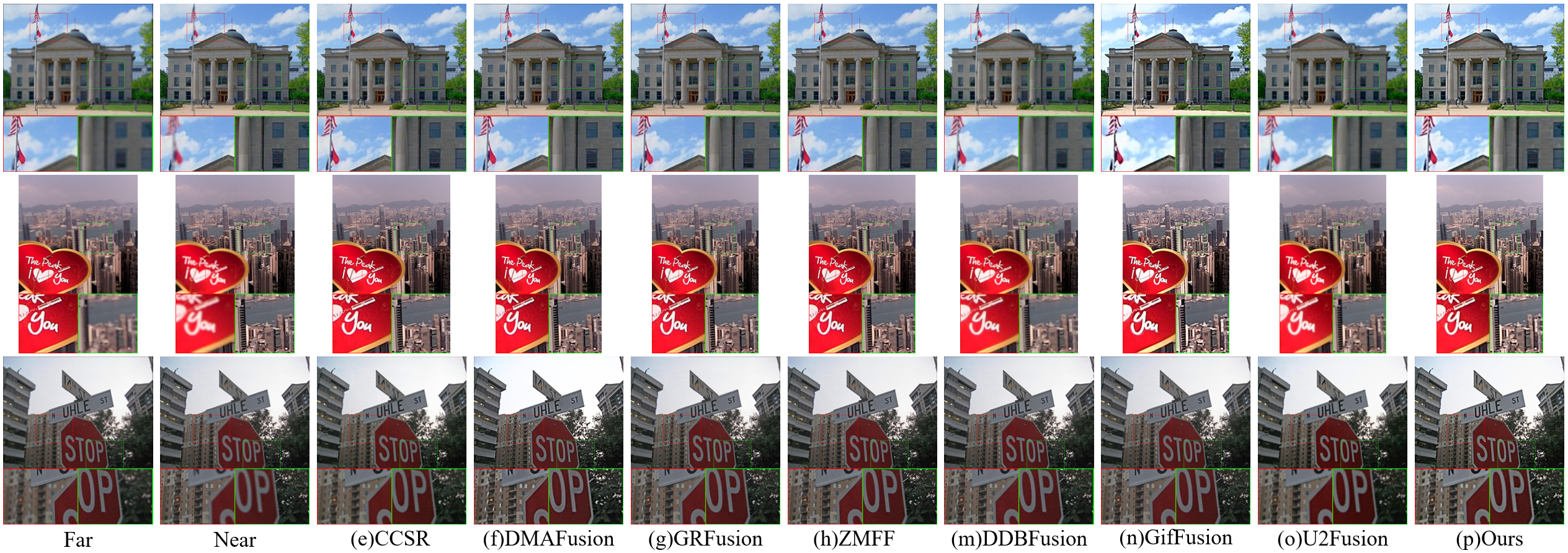}
	\caption{Qualitative comparison of fused results obtained by different methods on the MFF task.}
	\label{MFF}
\end{figure*}

\begin{table*}[t]
	\centering
	\caption{Quantitative comparison on Lytro and MFI-WHU datasets.}
	\label{MFF_lab}
	\resizebox{\textwidth}{!}{
		\begin{tabular}{lccccc lccccc}
			\toprule
			\multirow{2}{*}{Method} &
			\multicolumn{5}{c}{Lytro} &
			\multirow{2}{*}{Method} &
			\multicolumn{5}{c}{MFI-WHU} \\
			\cmidrule(lr){2-6} \cmidrule(lr){8-12}
			& EN & SD & VIF & SCD & CC
			&  & EN & SD & VIF & SCD & CC \\
			\midrule
			CCSR-Net
			& 7.531 & 57.565 & \underline{0.679} & 0.840 & \underline{0.971}
			& CCSR-Net & 7.319 & 52.379 & \underline{0.683} & 0.868 & \underline{0.967} \\
			
			DMANet
			& 7.531 & 57.569 & 0.451 & 0.762 & 0.838
			& DMANet & 7.317 & 52.429 & 0.420 & 0.770 & 0.816 \\
			
			GRFusion
			& 7.532 & 57.616 & \textbf{0.682} & \underline{0.841} & 0.970
			& GRFusion & 7.322 & 52.420 & \textbf{0.690} & \underline{0.872} & 0.965 \\
			
			ZMFF
			& 7.524 & 56.954 & 0.448 & 0.762 & 0.837
			& ZMFF & 7.295 & 51.994 & 0.419 & 0.765 & 0.814 \\
			
			BBDFusion
			& 7.535 & 57.153 & 0.490 & 0.718 & 0.852
			& BBDFusion & 7.312 & 52.557 & 0.422 & 0.774 & 0.817 \\
			
			GIFNet
			& \underline{7.598} & \textbf{64.609} & 0.453 & 0.803 & 0.957
			& GIFNet & \textbf{7.442} & \textbf{61.735} & 0.423 & 0.832 & 0.947 \\
			
			U2Fusion
			& 7.460 & 58.390 & 0.492 & 0.710 & 0.849
			& U2Fusion & 7.226 & 53.326 & 0.422 & 0.766 & 0.815 \\
			
			\textbf{Ours}
			& \textbf{7.634} & \underline{63.173} & 0.533 & \textbf{0.842} & \textbf{0.974}
			& \textbf{Ours} & \underline{7.413} & \underline{59.004} & 0.522 & \textbf{0.874} & \textbf{0.968} \\
			\bottomrule
		\end{tabular}
	}
\end{table*}
For convenient comparison, we present the fusion results of different methods in Fig.\ref{MFF}, with enlarged views of detailed regions.
In multi-focus image fusion, our method demonstrates superior performance in both edge features and fine structural details compared with other approaches.
It also maintains the vividness of the fused image consistent with the original inputs, which aligns with the proposed inference logic.
Therefore, our method shows clear advantages in texture handling for multi-focus images.
As shown in Table \ref{MFF_lab}, our approach achieves the best or second-best performance in EN, SD, SCD, and CC across different datasets.
These results further verify the effectiveness of the proposed method.
\subsubsection{Infrared and visible image fusion}
\begin{figure*}[htbp]  
	\centering
	\includegraphics[width=0.9\linewidth]{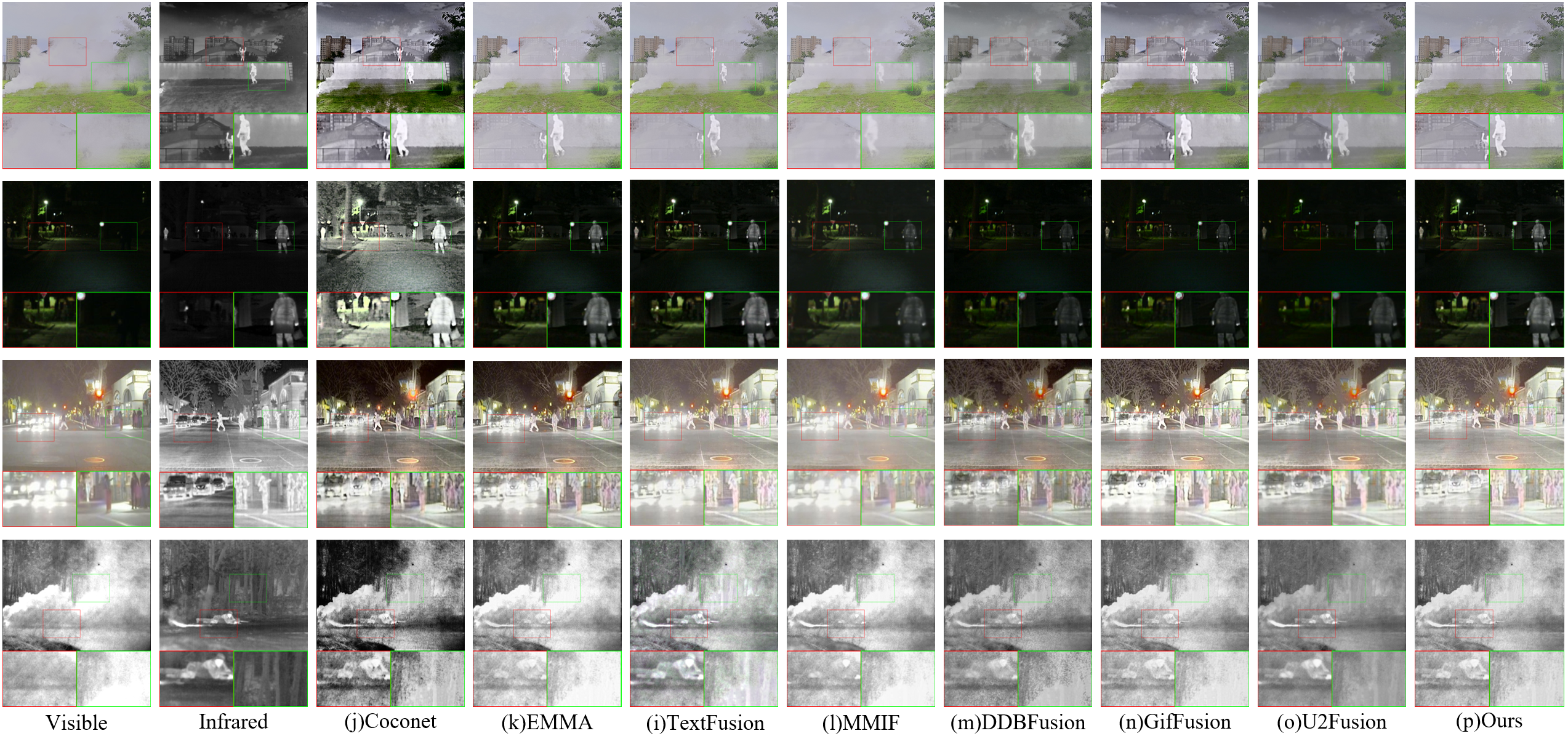}
	\caption{Qualitative comparison of fused results obtained by different methods on the VIF task.}
	\label{VIF}
\end{figure*}

\begin{table*}[t]
	\centering
	\caption{Quantitative comparison on M3FD and MSRS datasets.}
	\label{VIF_lab1}
	\resizebox{\textwidth}{!}{
		\begin{tabular}{lccccc lccccc}
			\toprule
			\multirow{2}{*}{Method} &
			\multicolumn{5}{c}{M3FD} &
			\multirow{2}{*}{Method} &
			\multicolumn{5}{c}{MSRS} \\
			\cmidrule(lr){2-6} \cmidrule(lr){8-12}
			& MI & PSNR & VIF & SCD & Qg
			&  & MI & PSNR & VIF & SCD & Qg \\
			\midrule
			TEXTFusion
			& \underline{3.638} & \textbf{16.390} & \textbf{0.379} & 0.731 & 0.833
			& TEXTFusion & 3.745 & 20.040 & 0.411 & 0.702 & 0.757 \\
			
			COCONet
			& 2.715 & 12.513 & 0.312 & 0.743 & 0.735
			& COCONet & 2.205 & 11.212 & 0.324 & 0.657 & 0.617 \\
			
			EMMA
			& 3.621 & \underline{16.375} & 0.321 & 0.767 & 0.837
			& EMMA & \underline{3.952} & \textbf{20.789} & 0.457 & \underline{0.740} & 0.859 \\
			
			MMIF-INet
			& 2.739 & 16.057 & 0.322 & \underline{0.783} & \underline{0.852}
			& MMIF-INet & 2.961 & 19.989 & \textbf{0.491} & 0.747 & \underline{0.862} \\
			
			BBDFusion
			& 1.136 & 12.345 & 0.018 & 0.321 & 0.525
			& BBDFusion & 2.171 & 18.962 & 0.300 & 0.731 & 0.818 \\
			
			GIFNet
			& 2.752 & 14.949 & 0.251 & 0.759 & 0.804
			& GIFNet & 1.969 & 18.518 & 0.289 & 0.690 & 0.806 \\
			
			U2Fusion
			& 2.818 & 15.591 & 0.251 & 0.781 & 0.782
			& U2Fusion & 2.146 & 18.745 & 0.249 & 0.730 & 0.685 \\
			
			\textbf{Ours}
			& \textbf{3.778} & 16.077 & \underline{0.324} & \textbf{0.785} & \textbf{0.857}
			& \textbf{Ours} & \textbf{3.997} & \underline{20.131} & \underline{0.458} & \textbf{0.752} & \textbf{0.869} \\
			\bottomrule
		\end{tabular}
	}
\end{table*}
\begin{table*}[t]
	\centering
	\caption{Quantitative comparison on RoadScene and TNO datasets.}
	\label{VIF_lab2}
	\resizebox{\textwidth}{!}{
		\begin{tabular}{lccccc lccccc}
			\toprule
			\multirow{2}{*}{Method} &
			\multicolumn{5}{c}{RoadScene} &
			\multirow{2}{*}{Method} &
			\multicolumn{5}{c}{TNO} \\
			\cmidrule(lr){2-6} \cmidrule(lr){8-12}
			& MI & PSNR & VIF & SCD & Qg
			&  & MI & PSNR & VIF & SCD & Qg \\
			\midrule
			TEXTFusion
			& \underline{3.537} & \underline{16.378} & \textbf{0.373} & 0.778 & 0.682
			& TEXTFusion & \underline{3.428} & \underline{15.464} & \underline{0.418} & 0.770 & 0.821 \\
			
			COCONet
			& 2.479 & 12.567 & 0.305 & 0.753 & 0.771
			& COCONet & 2.085 & 12.153 & 0.349 & 0.744 & 0.703 \\
			
			EMMA
			& 3.291 & 14.626 & \underline{0.360} & \underline{0.790} & 0.812
			& EMMA & 3.038 & 14.815 & 0.397 & 0.788 & 0.839 \\
			
			MMIF-INet
			& 2.626 & 16.206 & 0.348 & 0.785 & \underline{0.814}
			& MMIF-INet & 2.407 & 15.420 & 0.392 & 0.789 & \underline{0.855} \\
			
			BBDFusion
			& 2.680 & 14.789 & 0.291 & 0.781 & 0.814
			& BBDFusion & 2.034 & 15.065 & 0.338 & 0.770 & 0.826 \\
			
			GIFNet
			& 2.602 & 15.402 & 0.269 & 0.751 & 0.798
			& GIFNet & 2.031 & 15.397 & 0.278 & 0.751 & 0.812 \\
			
			U2Fusion
			& 3.025 & \textbf{16.429} & 0.305 & 0.774 & 0.800
			& U2Fusion & 2.110 & \textbf{15.901} & 0.292 & \underline{0.792} & 0.804 \\
			
			\textbf{Ours}
			& \textbf{3.610} & 16.016 & 0.337 & \textbf{0.791} & \textbf{0.826}
			& \textbf{Ours} & \textbf{3.630} & 15.168 & \textbf{0.422} & \textbf{0.794} & \textbf{0.867} \\
			\bottomrule
		\end{tabular}
	}
\end{table*}
Fig.\ref{VIF} presents the fusion results produced by different methods. 
For clearer visual inspection, the regions with rich details are further enlarged.
The figure illustrates the fusion performance under various challenging conditions, 
including smoke, nighttime scenes, and lighting interference.
Compared with other approaches, our method effectively preserves infrared features,
which benefits from the GBPC-based prior that establishes the luminance foundation 
of the fused image, making the infrared characteristics more prominent.
It also exhibits strong edge responses, contributing to robust performance across 
diverse environments.
As shown in Table~\ref{VIF_lab1} and Table~\ref{VIF_lab2}, our method achieves the best 
results in MI, SCD, and $Q_g$, and obtains high or second-best scores on the remaining 
metrics.
These results indicate that the proposed method can effectively preserve the edge 
structures and the original information in the fused images.

Although the proposed method adopts a few-shot learning strategy, it remains competitive when compared quantitatively and qualitatively with state-of-the-art approaches.
This validates the effectiveness of our method.
Under the guidance of the prior, the known prior information drives the neural network to extract complete fusion information from the source images, enabling the network to learn the entire fusion rule during this process.
Consequently, the method avoids relying on large-scale datasets for gradual rule extraction and reduces the dependence of the neural network on the training data.

\subsection{Efficiency Analysis}
We conducted experiments to evaluate the parameter size, FLOPs, and fusion time of different models.
All experiments were performed using input images of size $224\times224$, and each model was executed 100 times, with the average value taken as the final result.
As shown in Table \ref{method}, our method achieves the best performance in parameter count, FLOPs, and fusion time.
In our framework, the neural network infers the edge features of the true fused image based on an incomplete but known prior.
Therefore, the network architecture remains lightweight and is implemented as a clean CNN without requiring additional regularization structures to assist the inference process.
In contrast, other methods, such as diffusion-based models or deep network architectures, incur significantly higher computational costs.
These results demonstrate that the proposed method has strong deployability and practical efficiency advantages.

\section{Ablation Study and Theoretical Analysis}
\subsection{Incomplete prior and few-shot learning}
\begin{figure}[h]  
	\centering
	\includegraphics[width=0.85\linewidth]{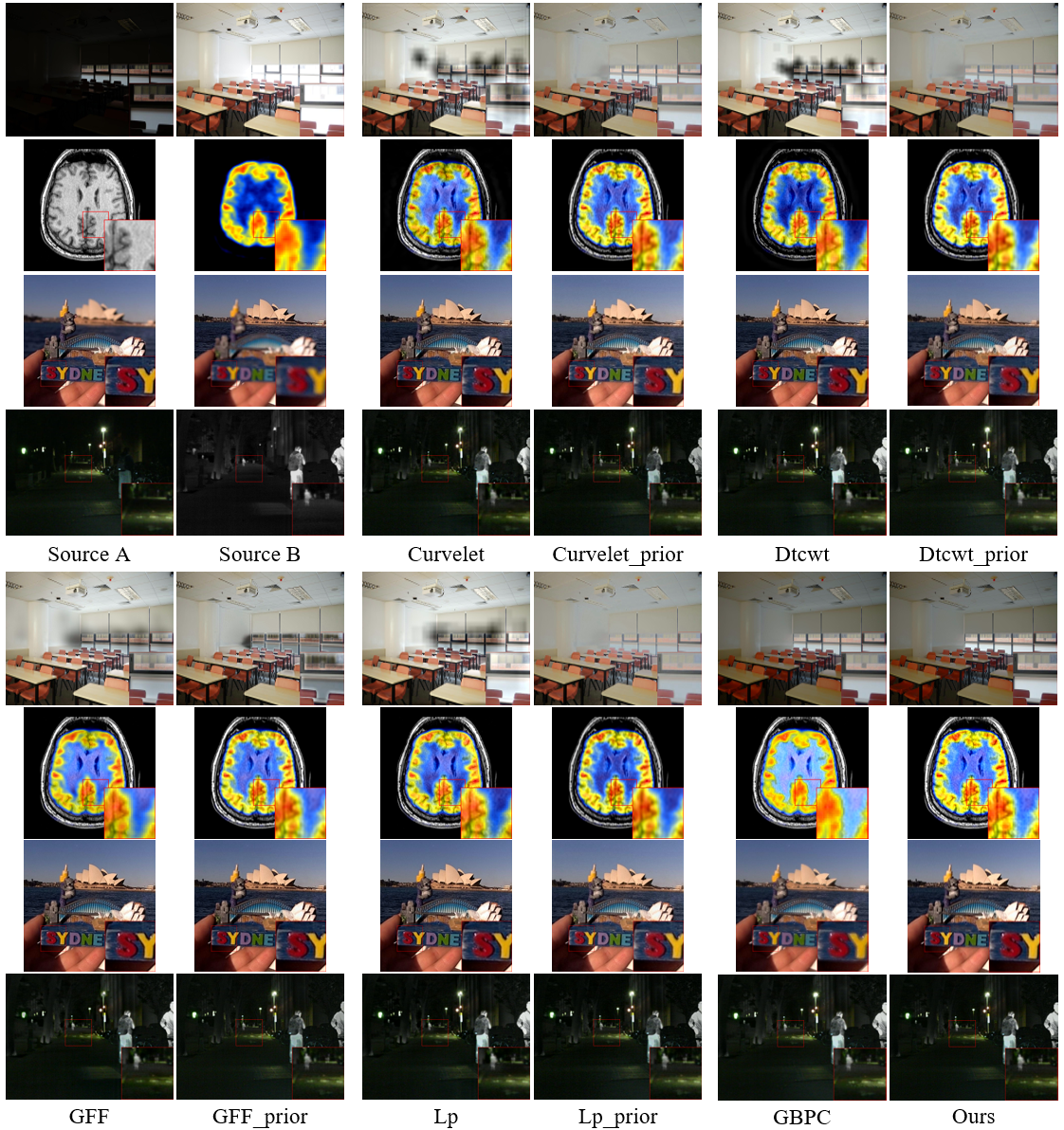}
	\caption{Visualization of fused results produced by different algorithms, along with the fused images obtained when these results are further used as priors for the neural network.}
	\label{prior}
\end{figure}
This paper proposes an incomplete prior and characterizes the deterministic and uncertain information within the algorithmic prior through the coarse-grained granular-ball similarity classes formed in the final stage. A loss function is further designed to establish the relationship between these two types of information. The incomplete prior generated by the Granular Ball Pixel Computation (GBPC) algorithm embodies an inference mechanism: in reliable regions, the model trusts the structural cues provided by the prior, while in uncertain regions, the network re-infers ambiguous edges and details from the source modalities rather than directly imitating the prior result. In this way, the overfitting problem caused by complete priors can be effectively avoided. In the few-shot learning scenario, we further exploit the incompleteness of the prior together with the self-recursive property introduced by image cropping to simulate complex real-world environments. Through this strategy, the network can adaptively adjust the learned fusion rules from limited samples, thereby enabling the inference of realistic fused images.

In contrast, traditional fusion algorithms generate complete priors. Such methods usually impose strong algorithmic preferences and provide a fully-constructed fusion image to the network. When a complete prior is injected into the model, it becomes difficult for the network to distinguish informative cues from algorithm-induced biases, leading the network to inherit the shortcomings of the prior during training and eventually weakening the fusion performance.
To validate the effectiveness of loss propagation and re-inference driven by an incomplete prior, we replace the GBPC prior with four representative complete priors. Specifically, we adopt: Curvelet (based on the curvelet transform), DTCWT (utilizing the dual-tree complex wavelet transform), GFF (using guided filtering), and LP (based on Laplacian pyramid decomposition). Since these priors do not provide an uncertainty ratio, we uniformly set their loss parameters as \( r_{\mathrm{POS}} = r_{\mathrm{BND}} = 0.5 \) and train them under identical small-sample settings. As shown in Fig.~\ref{prior} and Table~\ref{prior_lab}, networks guided by complete priors inherit method-specific artifacts, such as dark overstaining in MEF, even when the network attempts to refine edges from the source images. In contrast, the GBPC prior keeps uncertain regions through the BND and allows the network to continue reasoning instead of committing to an erroneous fusion pattern. Consequently, our approach achieves superior qualitative and quantitative results.
\begin{table*}[t]
	\centering
	\caption{Performance comparison of traditional priors, GBPC, and the proposed method across MEF, MIF, MFF, and VIF tasks.}
	\renewcommand{\arraystretch}{1.15}
	\label{prior_lab}
	\resizebox{\textwidth}{!}{
		\begin{tabular}{c|ccccc|ccccc|ccccc|ccccc}
			\hline
			\textbf{Method} 
			& \multicolumn{5}{c|}{\textbf{MEF}} 
			& \multicolumn{5}{c|}{\textbf{MIF}}
			& \multicolumn{5}{c|}{\textbf{MFF}}
			& \multicolumn{5}{c}{\textbf{VIF}} \\
			\hline
			& EN & MI & PSNR & CC & Qab
			& MI & PSNR & VIF & SCD & Qg
			& MI & SD & PSNR & AG & Qab
			& EN & SD & VIF & SCD & CC \\
			\hline
			
			Curvelet 
			& 6.491 & 3.995 & 8.500 & 0.459 & \underline{0.697}
			& 3.316 & 15.950 & 0.255 & 0.695 & 0.808
			& \underline{2.975} & 77.412 & 15.349 & 18.461 & 0.676
			& 7.098 & 59.972 & 0.451 & 0.758 & 0.896 \\
			
			Curvelet\_prior
			& 5.627 & 3.758 & 9.272 & \underline{0.673 }& 0.346
			& 3.209 & 16.022 & 0.237 & 0.715 & 0.814
			& 2.695 & 79.614 & 14.217 & 17.923 & 0.656
			& 7.130 & 58.340 & 0.404 & 0.804 & 0.934 \\
			
			Dtcwt
			& 6.448 & \underline{5.006} & 7.974 & 0.497 & 0.667
			& 3.278 & 15.486 & 0.240 & 0.704 & 0.809
			& 2.840 & 77.450 & \underline{15.309} & 18.513 & 0.689
			& 7.141 & 59.647 & 0.500 & 0.763 & 0.889 \\
			
			Dtcwt\_prior
			& 5.463 & 3.475 & 9.013 & 0.651 & 0.285
			& 3.129 & 15.659 & 0.227 & 0.720 & 0.811
			& 2.513 & 74.507 & 14.268 & 17.026 & 0.650
			&\underline{7.284 }& 58.385 & 0.470 & \underline{0.806} & 0.926 \\
			
			GFF
			& 6.316 & 3.739 & 8.197 & 0.440 & 0.693
			& \underline{3.508 }& 16.169 & 0.375 & 0.689 & 0.809
			& 2.877 & 77.470 & 15.311 & 18.466 & \underline{0.690}
			& 6.983 & 58.010 &\underline{ 0.510} & 0.764 & 0.890 \\
			
			GFF\_prior
			&\underline {6.843} & 3.588 & 8.158 & 0.626 & 0.676
			& 3.202 & 15.904 & 0.252 & 0.700 & 0.814
			& 2.640 & \underline{80.097 }& 14.677 & 19.266 & 0.676
			& 7.060 & 58.214 & 0.417 & 0.795 & \underline{0.949} \\
			
			Lp
			& 6.452 & 4.940 & 8.491 & 0.428 & 0.660
			& 2.549 & 15.455 & 0.297 & 0.713 & 0.796
			& 2.910 & 77.775 & 15.246 & 18.574 & 0.682
			& 7.156 & \underline{60.184} & 0.509 & 0.768 & 0.871 \\
			
			Lp\_prior
			& 5.343 & 3.600 & \underline{9.273} & 0.628 & 0.298
			& 3.133 & 15.838 & 0.237 & \underline{0.721} & 0.819
			& 2.590 & 77.167 & 13.946 & \underline{19.735} & 0.660
			& 7.245 & 59.422 & 0.460 & 0.790 & 0.890 \\
			
			GBPC
			& 6.636 & 4.495 & 8.370 & 0.582 & 0.476
			& 3.690 & \underline{16.407 }& \underline{0.377} & 0.671 & \underline{0.828}
			& 2.203 & 74.628 & 13.015 & 16.393 & 0.584
			& 6.870 & 58.863 & 0.443 & 0.743 & 0.921 \\
			
			\textbf{OURS}
			& \textbf{6.938} & \textbf{5.116} & \textbf{10.638} & \textbf{0.883} & \textbf{0.708}
			& \textbf{3.795} & \textbf{16.864} & \textbf{0.382} & \textbf{0.779} & \textbf{0.863}
			& \textbf{3.022} & \textbf{85.355} & \textbf{16.010} & \textbf{20.282} & \textbf{0.717}
			& \textbf{7.524} & \textbf{61.088} & \textbf{0.528} & \textbf{0.858} & \textbf{0.971} \\
			\hline
			
		\end{tabular}
	}
\end{table*}

\subsection{Influence of prior confidence}
In the GBPC algorithm, we compute the confidence of the prior to guide the design of the loss function. 
To investigate how this computation influences the networks re-inference process, we perturb the output 
values of $r_{\text{pos}}$ and $r_{\text{bnd}}$. Specifically, we set up three groups of controlled experiments:
(1) $r_{\text{POS}} = 1,\; r_{\text{BND}} = 0$;
(2) $r_{\text{POS}} = 0.5,\; r_{\text{BND}} = 0.5$;
(3) $r_{\text{POS}} = 0,\; r_{\text{BND}} = 1$.
The experimental results are shown in Table~\ref{abla_lab}, where a quantitative evaluation of the fused images is provided.
It can be observed that, due to the networks insufficient response to ambiguous regions, the fusion performance 
across multiple tasks is inferior to that of the guided method. This validates the rationality of the confidence 
computed by GBPC, demonstrating that it effectively guides the network to extract edge features from the source images.

\subsection{Modality awareness}
\begin{figure}[h]  
	\centering
	\includegraphics[width=0.85\linewidth]{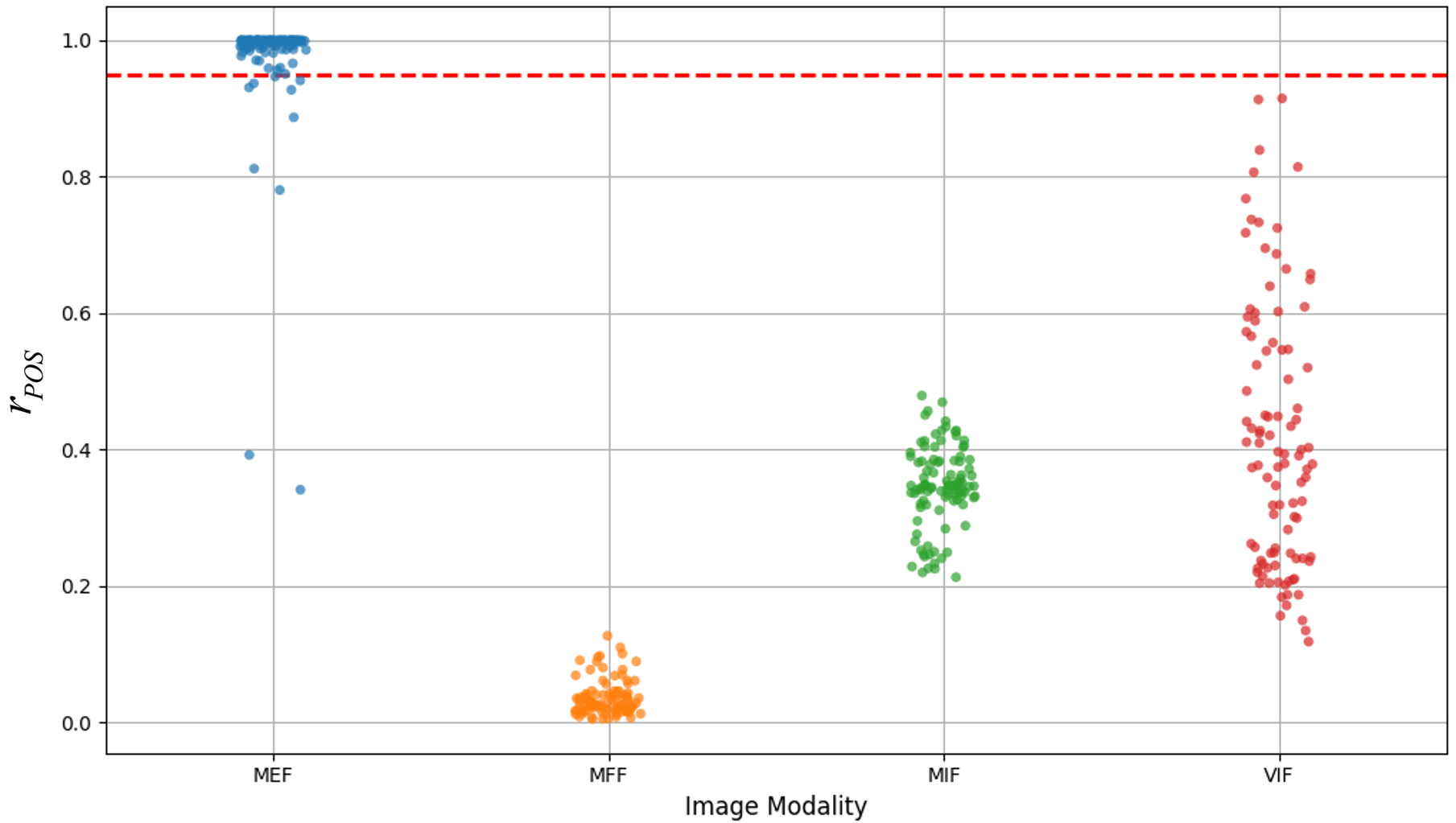}
	\caption{Distribution of $r_{\mathrm{POS}}$ measured by applying the GBPC algorithm to 100 randomly selected images from each modality.The red dashed line represents 0.95.}
	\label{class}
\end{figure}
In the GBPC algorithm, high-exposure regions usually exhibit significant intensity discrepancies, which cause more meta-granular balls to fall into the POS domain. As a result, the value of $r_{\mathrm{POS}}$ becomes significantly different from that observed in other modalities.

To verify this property, we analyze the statistical distribution of $r_{\mathrm{POS}}$ across different modalities. As shown in Fig.~\ref{class}, we randomly sample 100 image pairs from multiple datasets and compute the distribution of $r_{\mathrm{POS}}$. The results show that when the threshold is set to $r_{\mathrm{POS}} = 0.95$, high-exposure images can be clearly distinguished from other images.
Based on this observation, the threshold $M$ in the GBPC algorithm is set to $0.95$ as a fixed parameter to enable modality-aware perception. Within the proposed framework, cropped high-exposure images produce patches with varying exposure characteristics. After processing by GBPC, over-exposed regions are suppressed while structurally similar regions are enhanced, thereby guiding the neural network to learn appropriate fusion behaviors.
In the subsequent ablation study (Table~\ref{abla_lab}), we further conduct quantitative analysis. Since the modality perception mechanism is specifically designed to suppress over-exposed regions in multi-exposure fusion (MEF), it does not apply to other fusion modalities. Therefore, all ablation experiments related to modality perception are conducted only on the MEF task.
The results show that priors without modality awareness fail to adequately suppress high-exposure regions and lead to inferior performance in objective evaluation metrics.

\subsection{Loss function ablation}
In our framework, the prior is connected through the loss function, and the loss terms are adjusted to 
encourage the network to extract edge information from the source images. In this section, we conduct an 
ablation study on the loss function, specifically including: 
(1) \textbf{w/o $L_{POS}$}: removing the extraction of edge information from the prior; 
(2) \textbf{w/o $L_{BND}$ }: removing the extraction of edge features from the source images; 
(3) \textbf{w/o Laplacian}: removing the second-order derivative used for auxiliary inference. 
The quantitative results are shown in Table~\ref{abla_lab}. It can be observed that, when the source-image edge features 
are removed, the fused images exhibit degraded performance. When the prior-image edge information is omitted, 
the network lacks part of the necessary structural cues, resulting in incomplete information and suboptimal 
fusion outcomes. In contrast, incorporating the second-order derivative significantly improves the image quality, 
which validates the rationality and effectiveness of the designed loss function.

\begin{table*}[t]
	\centering
	\caption{Ablation results across MEF, MIF, MFF, and VIF tasks.}
	\renewcommand{\arraystretch}{1.15}
	\label{abla_lab}
	\resizebox{\textwidth}{!}{
		\begin{tabular}{c|ccccc|ccccc|ccccc|ccccc}
			\hline
			\textbf{Method} 
			& \multicolumn{5}{c|}{\textbf{MEF}} 
			& \multicolumn{5}{c|}{\textbf{MIF}}
			& \multicolumn{5}{c|}{\textbf{MFF}}
			& \multicolumn{5}{c}{\textbf{VIF}} \\
			\hline
			& EN & MI & PSNR & CC & Qab
			& MI & PSNR & VIF & SCD & Qg
			& MI & SD & PSNR & AG & Qab
			& EN & SD & VIF & SCD & CC \\
			\hline
			
			$r_{\text{POS}} = 0.5,\; r_{\text{BND}} = 0.5$
			& \underline{6.470 }& 4.925 & 9.858 & 0.872 & 0.632
			& \underline{3.772} & 16.051 & 0.340 & 0.720 & 0.860
			& 2.953 & 79.885 & 15.665 & \underline{18.719} & 0.668
			& 7.405 & 57.684 & 0.515 & \underline{0.850} & \underline{0.961} \\
			
			$r_{\text{POS}} = 0,\; r_{\text{BND}} = 1$
			& 6.545 & 4.234 & 9.871 & 0.861 & 0.680
			& 3.688 & 15.936 & 0.319 & 0.627 & 0.853
			& 2.884 & 81.075 & 15.417 & 18.600 & \underline{0.705}
			& \underline{7.466} & 59.763 & 0.512 & 0.837 & 0.954 \\
			
			$r_{\text{POS}} = 1,\; r_{\text{BND}} = 0$
			& 6.390 & \underline{5.048} & 9.027 & \underline{0.875} & 0.473
			& 3.704 & \underline{16.418} & \underline{0.339} & 0.716 & 0.853
			& \underline{3.005} & 73.393 & 15.895 & 14.674 & 0.566
			& 7.394 & 58.019 & \underline{0.519} & 0.838 & 0.959 \\
			\hline			 			
			w/o $L_{POS}$
			& 6.333 & 4.177 & 9.867 & 0.838 & 0.572
			& 3.638 & 16.162 & 0.319 & 0.720 & 0.855
			& 2.935 & 74.046 & \underline{15.929 }& 16.127 & 0.524
			& 7.436 & 60.060 & 0.497 & 0.834 & 0.955 \\
			
			w/o $L_{BND}$
			& 6.477 & 4.757 & 9.858 & 0.867 & 0.552
			& 3.626 & 16.337 & 0.269 & \underline{0.728} & 0.854
			& 2.910 & 74.783 & 15.027 & 16.136 & 0.523
			& 7.362 & 59.654 & 0.500 & 0.833 & 0.953 \\
			w/o $Laplacian$
			& 6.618 & 4.293 & \underline{9.922 }& 0.869 &\underline {0.701}
			& 3.616 & 15.767 & 0.331 & 0.727 &\underline{ 0.862}
			& 2.786 & \underline{81.334} & 15.820 & 18.864 & 0.701
			& 7.484 & \underline{60.977} & 0.508 & 0.842 & 0.954 \\
			\hline		
			w/o Modality Perception
			& 6.245 & 5.021 & 8.125 & 0.812 & 0.651
			& \multicolumn{5}{c|}{--}
			& \multicolumn{5}{c|}{--}
			& \multicolumn{5}{c}{--} \\
			
			\textbf{OURS}
			& \textbf{6.938} & \textbf{5.116} & \textbf{10.638} & \textbf{0.883} & \textbf{0.708}
			& \textbf{3.795} & \textbf{16.864} & \textbf{0.382} & \textbf{0.779} & \textbf{0.863}
			& \textbf{3.022} & \textbf{85.355} & \textbf{16.010} & \textbf{20.282} & \textbf{0.717}
			& \textbf{7.524} & \textbf{61.088} & \textbf{0.528} & \textbf{0.858} & \textbf{0.971} \\
			\hline
			
		\end{tabular}
	}
\end{table*}

\subsection{Influence of Sample Size}
\begin{figure}[h]  
	\centering
	\includegraphics[width=0.85\linewidth]{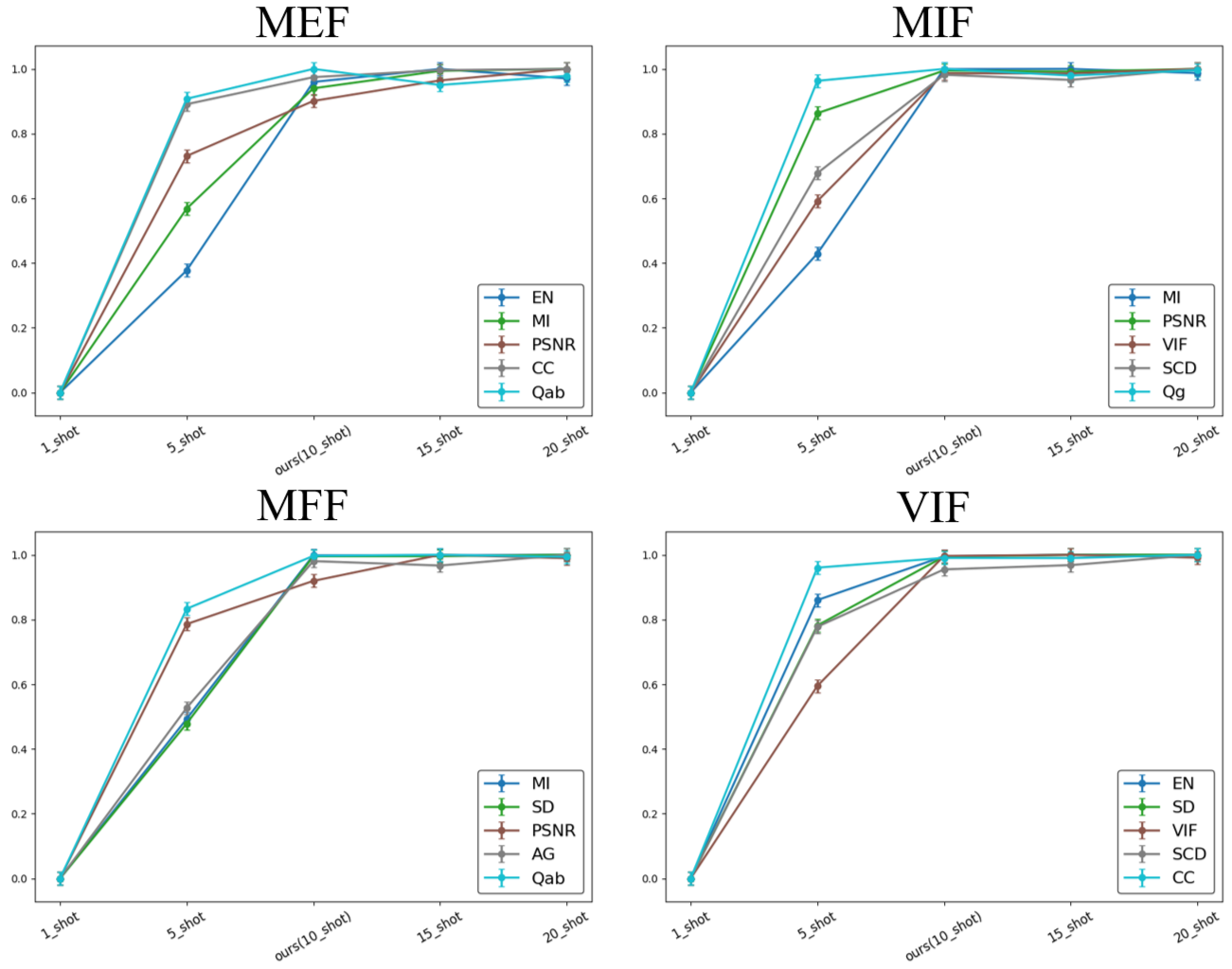}
	\caption{Quantitative analysis of fused images on different tasks under varying numbers of training samples.}
	\label{shot_dif}
\end{figure}

In this section, we investigate the influence of sample size on the performance 
of few-shot learning. We construct several comparison groups with different 
numbers of training samples, specifically: \textit{1\_shot}, \textit{5\_shot}, 
\textit{10\_shot (ours)}, \textit{15\_shot}, and \textit{20\_shot}. To ensure 
consistency across experiments, the selected samples follow an inclusion 
relationship, i.e., the 15\_shot subset is contained within the 20\_shot set. 
For each configuration, multiple random sampling trials are conducted, and the 
final results are obtained by averaging the objective evaluation metrics.
All experimental settings, including the number of iterations, remain identical, 
and the sample size serves as the only varying factor. As illustrated in Fig.~\ref{shot_dif}, 
all metrics are normalized to the range $(0,1)$ for clearer visualization. 
The results show that when the sample size is extremely limited, such as in 
the \textit{1\_shot} and \textit{5\_shot} settings, the generalization capability 
is significantly weakened. When the sample size exceeds 10, the objective metrics 
begin to stabilize, though mild fluctuations are still observed. This indicates 
that although our framework incorporates rule-based priors, the neural network 
still requires a minimum amount of data to effectively learn and enhance 
generalization.
Moreover, in our framework, the neural network must not only utilize the provided 
prior but also infer additional edge features from the source images. This 
necessitates a minimal number of training samples to ensure that the model can 
learn robust fusion rules and simulate real-world scenarios even from limited 
image fragments.

\subsection{Limitations}
In the proposed method, we expect the neural network to continue reasoning based on the available prior, thereby compensating for the incompleteness of the prior. Although the experimental results demonstrate that the proposed method exhibits good generalization ability, several issues remain unresolved in the current experiments. Specifically:

\MakeUppercase{\romannumeral 1}  .The prior is provided by the GBPC algorithm, which is primarily suitable for typical image fusion tasks. For broader future applications, additional subspaces may be required to enable further inference.

\MakeUppercase{\romannumeral 2}.The few-shot learning strategy in this work relies on image cropping to simulate various conditions, allowing the adaptive prior-generation algorithm to reconstruct fusion rules on cropped images. This design shifts the networks learning paradigm from extracting features solely from the dataset to re-inference based on the prior, thereby reducing reliance on large datasets. However, if the few-shot dataset contains overly extreme samples, the learned weights may exhibit reduced generalization capability.

\section{CONCLUSION}
In this work, we extend granular computing to general-purpose multimodal image fusion. Based on the principles of image fusion, we design the Granular Ball Pixel Computing (GBPC) algorithm and introduce the concept of an incomplete prior, thereby establishing a sample-level adaptive learning framework that enables few-shot training for neural networks.
Specifically, the GBPC algorithm performs multi-granularity analysis to extract informative structures from fused images. At the fine granularity, the equivalence principle is employed to compute pixel-level fusion weights. At the coarse granularity, fuzzy similarity is utilized to estimate the reliability of the prior, enabling adaptive regulation of the loss function during network training.
Experimental analyses demonstrate the advantages of the incomplete prior for neural network learning and verify the effectiveness of the proposed modality-awareness mechanism across different fusion tasks. Under a limited training setting with only ten samples, both qualitative and quantitative comparisons with state-of-the-art methods validate the effectiveness of the proposed framework.
Finally, we discuss the limitations of the proposed approach and investigate the influence of sample size on the generalization capability of the few-shot learning paradigm. In future work, we plan to incorporate additional subspaces to further enhance the accuracy and expressiveness of the model, and extend the framework to broader image processing tasks such as image segmentation and super-resolution.

\printcredits

\bibliographystyle{cas-model2-names}
\bibliography{my}
\end{document}